\documentclass{pasa}%
\usepackage{appendix}
\usepackage{supertabular}
\usepackage{graphics}
\usepackage{amsmath}
\usepackage{subfigure}
\usepackage{threeparttable}
\usepackage{multirow}
\usepackage{upgreek}

\newcommand{\mr}{\multirow}

\usepackage{longtable}
\usepackage[tableposition=below]{caption}
\usepackage{rotating}
\usepackage{array}
\newcolumntype{L}{>{\arraybackslash}m{\columnwidth}}

\newcommand*{\affaddr}[1]{#1} 
\newcommand*{\affmark}[1][*]{\textsuperscript{#1}}

\title[Follow up of GW170817 and its electromagnetic counterpart by Australian-led observing programs]{Follow up of GW170817 and its electromagnetic counterpart by Australian-led observing programs}

\author[Andreoni et al.]{
I.~Andreoni$^{1,2,3}$\thanks{igor.andreoni@gmail.com},
K.~Ackley$^{2,4,5}$,
J.~Cooke$^{1,2,6}$,
A.~Acharyya$^{7}$,
J.~R.~Allison$^{8,9}$,
G.~E.~Anderson$^{10}$,
M.~C.~B.~Ashley$^{11}$,
D.~Baade$^{12}$,
M.~Bailes$^{1,2}$,
K.~Bannister$^{13}$
% ...
}

\jid{PASA}
\doi{10.1017/pas.\the\year.xxx}
\jyear{\the\year}

\usepackage[authoryear]{natbib}
\bibpunct{(}{)}{;}{a}{}{,}
\usepackage{aas_macros}
\usepackage{hyperref} 
\hypersetup{colorlinks,citecolor=blue,linkcolor=blue,urlcolor=blue}

\begin{document}%
\begin{abstract}
The discovery of the first electromagnetic counterpart to a gravitational wave signal has generated follow-up observations by over 50 facilities world-wide, ushering in the new era of multi-messenger astronomy.  In this paper, we present follow-up observations of the gravitational wave event GW170817 and its electromagnetic counterpart SSS17a/DLT17ck (IAU label AT2017gfo) by 14 Australian telescopes and partner observatories as part of Australian-based and Australian-led research programs.  We report early- to late-time multi-wavelength observations, including optical imaging and spectroscopy, mid-infrared imaging, radio imaging, and searches for fast radio bursts.  Our optical spectra reveal that the transient source afterglow cooled from approximately 6400K to 2100K over a 7-day period and produced no significant optical emission lines.  The spectral profiles, cooling rate, and photometric light curves are consistent with the expected outburst and subsequent processes of a binary neutron star merger.  Star formation in the host galaxy probably ceased at least a Gyr ago, although there is evidence for a galaxy merger.  Binary pulsars with short (100 Myr) decay times are therefore unlikely progenitors, but pulsars like PSR B1534+12 with its 2.7 Gyr coalescence time could produce such a merger.  The displacement ($\sim$2.2 kpc) of the binary star system from the centre of the main galaxy is not unusual for stars in the host galaxy or stars originating in the merging galaxy, and therefore any constraints on the kick velocity imparted to the progenitor are poor.
\end{abstract}
\maketitle%
%
\onecolumn
\section*{AFFILIATIONS }
\label{sec:affiliations}
\affaddr{\affmark[1]Centre for Astrophysics and Supercomputing, Swinburne University of Technology, PO Box 218, H29, Hawthorn, VIC 3122, Australia}\\
\affaddr{\affmark[2]The Australian Research Council Centre of Excellence for Gravitational Wave Discovery (OzGrav)} \\
\affaddr{\affmark[3]Australian Astronomical Observatory, 105 Delhi Rd, North Ryde, NSW 2113, Australia}\\
\affaddr{\affmark[4]Monash Centre of Astrophysics, Monash University, VIC 3800, Australia}\\ 
\affaddr{\affmark[5]School of Physics \& Astronomy, Monash University, VIC 3800, Australia}\\
\affaddr{\affmark[6]The Australian Research Council Centre of Excellence for All-Sky Astrophysics (CAASTRO)}\\
\affaddr{\affmark[7]Research School of Astronomy and Astrophysics, The Australian National University, Canberra ACT 2611, Australia}\\
\affaddr{\affmark[8]Sydney Institute for Astronomy, School of Physics, University of Sydney, NSW 2006, Australia}\\
\affaddr{\affmark[9]The Australian Research Council Centre of Excellence for All-sky Astrophysics in 3 Dimensions (ASTRO 3D)}\\
\affaddr{\affmark[10]International Centre for Radio Astronomy Research, Curtin University, Bentley WA 6102,  Australia}\\
\affaddr{\affmark[11]School of Physics, University of New South Wales, NSW 2052, Australia}\\
\affaddr{\affmark[12]European Organisation for Astronomical Research in the Southern Hemisphere (ESO), Karl-Schwarzschild-Str.~2, 85748 Garching bei M{\"u}nchen, Germany}\\
\affaddr{\affmark[13]ATNF, CSIRO Astronomy and Space Science, PO Box 76, Epping, NSW 1710, Australia}\\
\affaddr{\affmark[14]School of Earth and Space Exploration, Arizona State University, Tempe, AZ 85287, USA}\\
\affaddr{\affmark[15]Department of Applied Geology, Curtin University, GPO Box U1987, Perth, WA 6845, Australia}\\
\affaddr{\affmark[16]Art{\'e}mis/Observatoire de la Cote d'Azur/CNRS, Boulevard de l'Observatoire CS 34229 - F 06304 NICE Cedex 4, France}\\
\affaddr{\affmark[17]Department of Astronomy, Stockholm University, Albanova, SE 10691 Stockholm, Sweden}\\
\affaddr{\affmark[18]Department of Physics, University of Wisconsin--Milwaukee, Milwaukee, WI 53201, USA}\\
\affaddr{\affmark[19]South African Astronomical Observatory, PO Box 9, 7935 Observatory, South Africa}\\
\affaddr{\affmark[20]School of Physics, University of Western Australia, Crawley, WA 6009, Australia}\\
\affaddr{\affmark[21]University of the Virgin Islands, 2 John Brewer's Bay, 00802 St Thomas, US Virgin Islands, USA}\\
\affaddr{\affmark[22]School of Physics and Astronomy, University of Nottingham, Nottingham, UK}\\
\affaddr{\affmark[23]ATNF, CSIRO Astronomy and Space Science, 26 Dick Perry Avenue, Kensington WA 6152, Australia}\\
\affaddr{\affmark[24]Dunlap Institute for Astronomy and Astrophysics, University of Toronto, ON, M5S 3H4, Canada}\\
\affaddr{\affmark[25]Department of Physics, 2354 Fairchild Drive, U.S. Air Force Academy, CO 80840, USA}\\
\affaddr{\affmark[26]The Oskar Klein Centre, Department of Physics, Stockholm University, Albanova, SE 106 91 Stockholm, Sweden } \\
\affaddr{\affmark[27]Department of Mechanical Engineering, Curtin University, GPO Box U1987, Perth, WA 6845, Australia}\\
\affaddr{\affmark[28]Purple Mountain Observatory, Chinese Academy of Sciences, Nanjing 210008, China}\\
\affaddr{\affmark[29]Chinese Center for Antarctic Astronomy, Nanjing 210008, China}\\
\affaddr{\affmark[30]National Astronomical Observatories, Chinese Academy of Sciences, Beijing 100012, China}\\
\affaddr{\affmark[31]Peripety Scientific Ltd., PO Box 11355 Manners Street, Wellington, 6142, New Zealand}\\
\affaddr{\affmark[32]Division of Physics, Mathematics and Astronomy, California Institute of Technology, Pasadena, CA 91125, USA}\\
\affaddr{\affmark[33]SKA Organisation, Jodrell Bank Observatory, SK11 9DL, UK}\\
\affaddr{\affmark[34]IRAP (CNRS/UPS) 34 ave. Edouard Belin, 31400 Toulouse France}\\
\affaddr{\affmark[35]Nanjing Institute of Astronomical Optics and Technology, Nanjing 210042, China}\\
\affaddr{\affmark[36]School of Astronomy and Space Science and Key Laboratory of Modern Astronomy and Astrophysics in Ministry of Education, Nanjing University, Nanjing 210093, China}\\
\affaddr{\affmark[37]Yunnan Observatories, Chinese Academy of Sciences, 650011 Kunming, Yunnan Province, China}\\
\affaddr{\affmark[38]Las Cumbres Observatory, 6740 Cortona Dr., Suite 102, Goleta, CA 93117-5575, USA}\\
\affaddr{\affmark[39]Department of Physics, University of California, Santa Barbara, CA 93106-9530, USA}\\
\affaddr{\affmark[40]Department of Physics, University of Washington, Seattle, WA 98195, USA}\\
\affaddr{\affmark[41]OrangeWave Innovative Science LLC, Moncks Corner, SC 29461, USA}\\
\affaddr{\affmark[42]Southern African Large Telescope Foundation, P.O. Box 9, 7935 Observatory, South Africa.}\\
\affaddr{\affmark[43]Centre for Translational Data Science, University of Sydney, NSW 2006, Australia}\\
\affaddr{\affmark[44]Centre de Recherche en Astronomie, Astrophysique et G\'eophysique, BP 63, Route de l'Observatoire, Bouzareah, 16340, Alger, Algeria}\\
\affaddr{\affmark[45]Tianjin Normal University, Tianjin 300074, China}\\
\affaddr{\affmark[46]Kavli Institute for Astronomy and Astrophysics, Peking University, 5 Yiheyuan Road, Haidian District, Beijing 100871, P.~R.~China}\\
\affaddr{\affmark[47]Department of Astrophysics, American Museum of Natural History, Central Park West and 79th Street, New York, NY 10024, USA}\\
\affaddr{\affmark[48]Institute of Astronomy, University of Cambridge, Madingley Road, Cambridge CB3 0HA, UK}\\
\affaddr{\affmark[49]George P. and Cynthia Woods Mitchell Institute for Fundamental Physics \& Astronomy, Texas A. \& M. University, Department of Physics and Astronomy, 4242 TAMU, College Station, TX 77843, USA}\\
\affaddr{\affmark[50]Aix Marseille Univ, CNRS/IN2P3, CPPM, Marseille, France}\\
\affaddr{\affmark[51]Physics Department and Tsinghua Center for Astrophysics, Tsinghua University, Beijing, 100084, China}\\
\affaddr{\affmark[52]Observatoire d'Auragne, 31190 Auragne, France}\\

\twocolumn

\section{INTRODUCTION }
\label{sec:intro}

The first detection of an electromagnetic (EM) counterpart to a gravitational wave (GW) event has led to the new era of gravitational-wave multi-messenger astrophysics.  The close coordination of LIGO data analysis groups and multiple observational teams worldwide via the restricted Gamma-Ray Coordinates Network (GCN) reports under confidential Memoranda of Understanding (MoU), were key to the prompt identification and detailed multi-wavelength follow up of the counterpart.

On August 17, 2017 12:41:04\footnote{All dates in this paper are \small{UT}, unless a different time reference is explicitly specified.} the Advanced Laser Interferometer Gravitational-Wave Observatory (aLIGO) interferometers detected a 
GW signal G298048, now referred to as GW170817 \citep[][]{GCN21505, GCN21509, GCN21510, GCN21513, GCN21527,LVCGW170817discovery}. The Advanced-Virgo (aVirgo) interferometer was online at the time of the discovery and also contributed to the localisation of the GW event. On August 17, 2017 12:41:06, about 2 seconds after the GW detection, the Gamma-ray Burst Monitor (GBM) instrument on board the {\it Fermi} satellite independently detected a short gamma-ray burst, labeled as GRB~170817A \citep[{\it Fermi}:][]{GCN21506, GCN21528, GCN21520,Goldstein2017}, \citep[INTEGRAL:][]{GCN21507, Savchenko2017}, \citep{LIGOandFermi2017}.  The close temporal coincidence of the gamma-ray burst and gravitational wave event made it a compelling target for follow up observations at other wavelengths. 

The One-Meter, Two-Hemisphere project (1M2H) first announced the discovery of a transient in an image acquired with the 1\,m Swope telescope at Las Campanas Observatory in Chile on 17 August 2017 at 23:33 UTC, 10.87\,hr after the LIGO detection.  However, the optical counterpart to GW170817 (and GRB~170817A) was already imaged independently by six other programs before this report.  The 1M2H team referred to the transient with the name Swope Supernova Survey 2017a \citep[SSS17a,][]{GCN21529, Coulter2017}.  Details about the other independent detections can be found in \cite{GCN21530} for the Dark Energy Camera; \cite{GCN21531} and \cite{Valenti2017} for the Distance Less Than 40\,Mpc survey (DLT40); \cite{GCN21538} and \cite{ArcaviPaper} for the Las Cumbres Observatory; \cite{GCN21544} and \cite{Tanvir2017} for the Visible and Infrared Survey Telescope for Astronomy; and \cite{GCN21546} and Lipunov et al. (2017, in prep) for the MASTER discoveries.

\cite{LVCGW170817MMA} offer an extensive review of the world-wide follow up. The optical transient is located at RA = 13:09:48.089 DEC = --23:22:53.350 \citep{GCN21816,Kasliwal2017}, approximately 2.2 kpc from the centre of its host galaxy NGC~4993.  The host is a nearby E/S0 galaxy at z = 0.009727, corresponding to a distance of $\sim$39.5~Mpc \citep{Freedman2001}.  Hereafter, we refer to the EM counterpart of GW170817 with the IAU label AT2017gfo. 

Short-duration GRBs (sGRBs) were previously suggested to be associated with merging compact objects, such as a binary neutron star (BNS) system or neutron star-black hole (NSBH) system \citep[e.g.,][]{Paczynski1986,Goodman1986}. Electromagnetically, such mergers are also postulated to generate a relatively rapidly evolving optical/infrared transient - referred to as kilonova or macronova \citep[e.g.,][]{Li1998, Barnes2013, Tanaka2013, Kasen2015, Metzger2015, Barnes2016}.  The combination of an sGRB and kilonova is considered the ``smoking gun'' signature of such mergers.  Kilonova candidates were previously identified during the follow up of sGRBs, for example GRB~080503 \citep{Perley2009,Gao2015},  GRB~130603B \citep{Berger2013k,Tanvir2013,Hotokezaka2013}, and GRB~050709 \citep{Jin2016}.  However, no kilonova candidates have been discovered unrelated to GRB triggers, despite their anticipated isotropic emission, unlike that of sGRBs. BNS and NSBH mergers, thus sGRBs, and subsequent kilonovae, are expected to be the most promising GW events to exhibit EM counterparts.

Previous work has discussed the importance of rapid response \citep[e.g.,][]{Chu2016} and collaborative strategies to maximise the chances of success in the EM follow up of aLIGO and Virgo triggers.  Specifically, \cite{Howell2015} presents the role that Australia can play in this context.  The association of GW170817 to GRB~170817A, detected during the LIGO and Virgo Collaboration (LVC) `O2' run, has enabled the first multi-messenger (EM multi-wavelength, neutrino, and GW observations) study of an astrophysical event \citep{LVCGW170817MMA}.
 
This paper presents and discusses the data acquired during the search for an EM counterpart to GW170817 and the follow up of the now confirmed counterpart, AT2017gfo, by 15 observing programs led by Australian institutions and researchers.  The observing programs include facilities and collaborators associated with the Australia Research Council (ARC) Centre of Excellence for All-sky Astrophysics (CAASTRO\footnote{\url{http://www.caastro.org}}), the ARC Centre of Excellence for Gravitational Wave Discovery (OzGrav\footnote{\url{http://www.ozgrav.org}}), and the multi-wavelength, multi-facility Deeper, Wider, Faster (DWF\footnote{\url{http://www.dwfprogram.altervista.org}}) program.   In Section~\ref{sec:instruments} we summarise the observations from the telescopes/instruments that participated in the GW170817 follow up, including optical, mid-infrared and radio imaging and spectroscopic observations.   In Section~\ref{sec:models} we provide an overview the spectroscopic observations of the event and host galaxy and preliminary fits of our observations to theoretical sGRB afterglow and kilonova models.  Finally, we present a discussion and summary and in Section~\ref{sec: conclusion}. 


\section{Facilities involved in the EM follow up of GW170817}
\label{sec:instruments}



The following sections describe the optical, mid-infrared (mIR), and radio telescopes, instruments, and relevant observations involved in the follow up of the GW170817 EM counterpart by Australian or Australian-led programs.  

\begin{figure*}
\centering
\includegraphics[width=2\columnwidth]{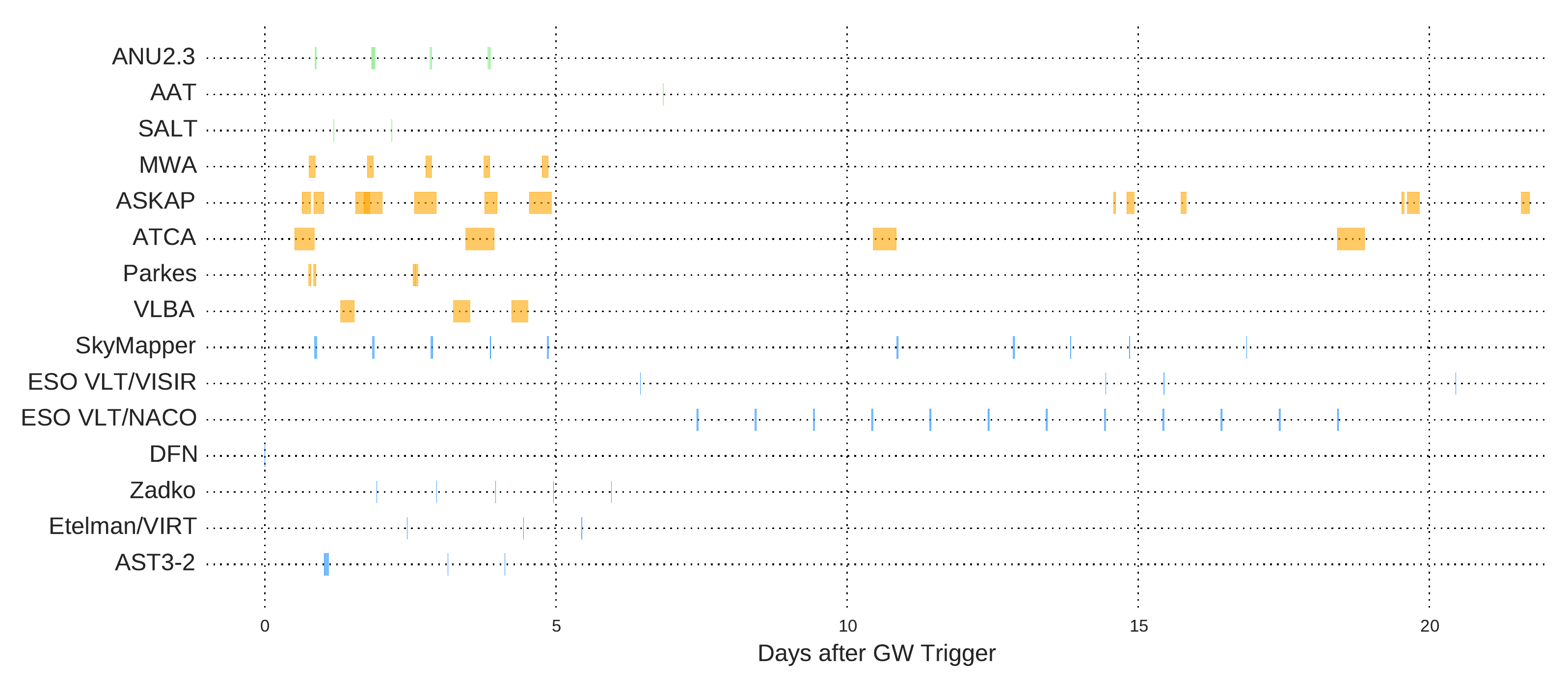}
\caption{Observation timeline for the facilities presented in this paper showing the time of observation offset from the GW event and the nominal length of the reported observations. Spectroscopic observations are shown in green, radio observations in orange, and optical and mid-infrared are in blue.}
\label{figure:timeline}
\end{figure*}

Shortly after the LVC community was alerted to the GW event, many of the facilities discussed here were triggered into action for follow-up observations.  However, NGC 4993 and the bulk of the LVC error ellipse had set in Eastern Australia and the Zadko telescope in Western Australia was temporarily not operational.  On the following day, the location  of the optical counterpart AT2017gfo was known.  Radio telescopes were on the field that day and optical facilities were on the field shortly after sunset.  Figure~\ref{figure:timeline} presents the broad temporal coverage of the GW event by our spectroscopic, radio, and optical/mIR observations that extend from early to late times.  The general characteristics of each facility is presented in Table~\ref{tab:instruments} and details of the corresponding observations are listed in Tables~\ref{table: obs and photometry AST3}--\ref{table: obs and photometry SALT}.

\begin{table*}
\centering
\begin{threeparttable}
\caption{Facilities participating in the follow up observations summarised in this paper.  Principal references for the relevant data from each facility are indicated in the right-most column.  We specify under which program the observations were taken when multiple groups used the same telescope to follow up GW170817 and AT2017gfo. } 
\label{tab:instruments}
\begin{tabular}{lllll}%
\hline
OIR Imaging& Band & FoV & Aperture (m) & References \\ 
\hline
SkyMapper & u,v,g,r,i,z & 5.7 deg$^2$ & 1.35 & this work \\ 
AST3-2\tnote{\textasteriskcentered} & i & 4.14 deg$^2$ & 0.5  &\small \cite{Hu2017}\\
Zadko & r, Clear & 0.15 deg$^2$ & 1 & this work \\  
UVI Etelman\tnote{\textasteriskcentered} & R, Clear & 0.11 deg$^2$ & 0.5 & this work \\
ESO VLT/NACO\tnote{\textasteriskcentered} & L3.8\,$\upmu$m & 784 arcsec$^2$ & 8.2 & this work \\
ESO VLT/VISIR\tnote{\textasteriskcentered} &  J8.9\,$\upmu$m & 1 arcmin$^2$ & 8.2 & \small{\cite{Kasliwal2017}}  \\
DFN & V & full-sky& 2$\times$8mm & this work \\
\hline
\hline
OIR Spectroscopy & Range (\AA) & R &Aperture (m) & References  \\ 
\hline
ANU2.3/WiFeS & 3300--9200 &  3000,7000& 2.3 & this work \\
\mr{3}{*}{SALT/RSS\tnote{\textasteriskcentered}} & \mr{3}{*}{3600--9700} & \mr{3}{*}{$\sim$300} & \mr{3}{*}{10} & this work \\
& & & & \small{\cite{McCully2017}} \\
& & & & \small{\cite{Buckley2017}} \\

AAT/2dF+AAOmega\tnote{\textasteriskcentered} & 3700--8800 & 1700 & 3.9 & this work \\
\hline
\hline
Radio  & Band & FoV (deg$^2$) & Mode & References \\ 
\hline
\mr{2}{*}{ATCA\tnote{\textdagger}} & \mr{2}{*}{5.5--21.2~GHz} & \mr{2}{*}{0.037 -- 0.143} & \mr{2}{*}{Imaging} & \small{\cite{Hallinan2017}}\\
 & & & & \small \cite{Kasliwal2017}\\
ASKAP & 0.7--1.8~GHz & 30  &Imaging & this work \\ 
ASKAP & 0.7--1.8~GHz & 210  &FRB & this work \\ 
MWA\tnote{\textsection} & 185\,MHz & 400 & Imaging &  this work  \\
VLBA & 8.7~GHz & 0.04 & Imaging & this work \\
Parkes & 1.2-1.6\,GHz & 0.55~deg$^2$ & FRB & this work \\
\hline
\end{tabular}
\begin{tablenotes}
\item [\textasteriskcentered] \footnotesize{Observations initiated, or proposed for, via collaboration with DWF program}
\item [\textdagger] \footnotesize{program CX391}
\item [\textdaggerdbl] \footnotesize{program BD218}
\item [\textsection] \footnotesize{program D0010}
\end{tablenotes}
\end{threeparttable}
\end{table*}

\subsection{Optical/Near-Infrared Imaging}

\subsubsection{SkyMapper}

SkyMapper \citep{Keller2007} is a 1.35\,m modified-Cassegrain telescope located at Siding Spring Observatory in New South Wales, Australia, which is owned and operated by the Australian National University (ANU).  The camera has a 5.7 deg$^2$ field of view, a pixel scale of 0.5 arcsec/pixel and six photometric filters in the \textit{uvgriz} system, which span the visible and ultraviolet bands from 325\,nm to 960\,nm.  Typical single-epoch 5$\sigma$ limiting magnitudes for each filter are $u$=19.5, $v$=19.5, $g$=21, $r$=21, $i$=20, and $z$=19, over 100\,s exposure times.  Since 2014, SkyMapper has conducted a full-hemisphere Southern sky survey in all six bands \citep[see][and \url{http://skymapper.anu.edu.au}]{Wolf2017inprep}.  Alongside this survey, the SkyMapper Transient Survey (SMT) has been performing a survey dedicated to supernovae and other transients \citep{Scalzo2017}.

\begin{figure}
\begin{center}
\includegraphics[width=\columnwidth]{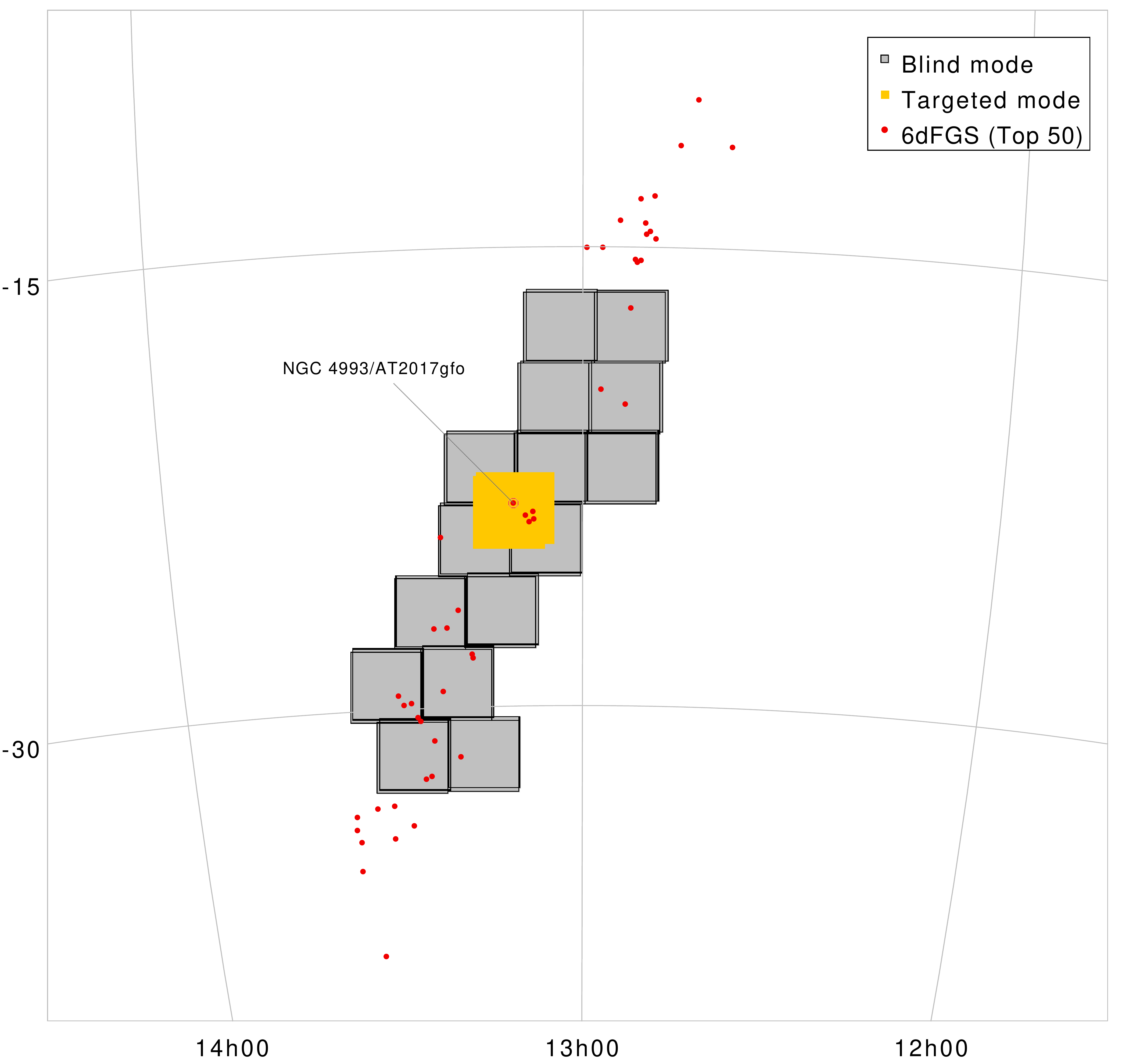}
\caption{Footprints of SkyMapper observations in two different follow-up modes: one using the blind search of new transient sources where fields overlap with GW localisation map (grey squares) and the other using the targeted observation of the optical counterpart, AT2017gfo, discovered by other EM follow-up groups (yellow square).  The positions of AT2017gfo and its host galaxy (NGC 4993) are indicated on the figure. The red dots are target galaxies from the 6dFGS catalogue that were prioritized by their position and spectroscopic redshift.}
\label{figure: SM fields}
\end{center}
\end{figure}

SkyMapper first received the GW trigger when the target area had recently set in Eastern Australia and began observing relevant target ranges shortly after sunset the following night.  The follow-up strategy included two components: (1) to obtain $uvgriz$ photometry of the field containing AT2017gfo, in the event that the transient was the correct counterpart to the GW trigger, and (2) to image the 90\% probability region (85\,deg$^2$) of the LVC sky-map to search for other counterpart candidates (Figure~\ref{figure: SM fields}).

\begin{figure}[!ht]
\centering
\includegraphics[width=\columnwidth]{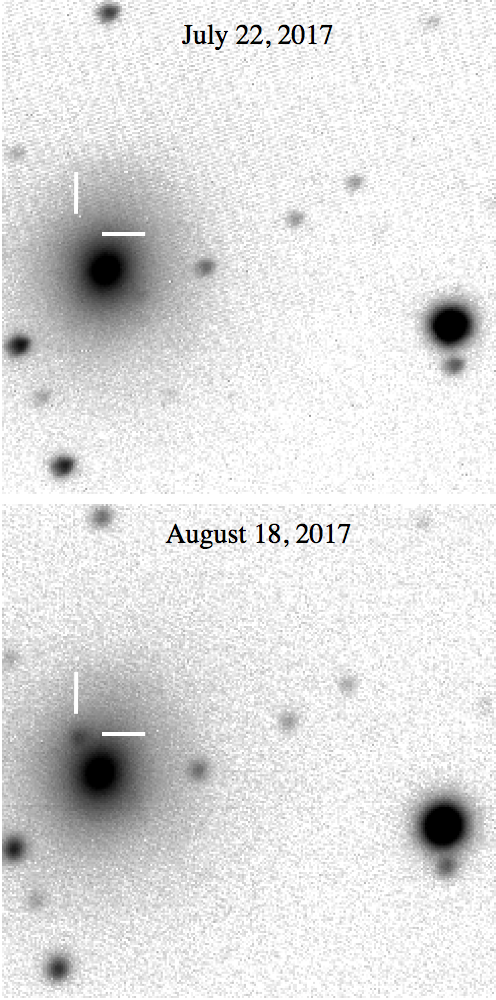}
\caption{SkyMapper optical images of NGC~4993 (left centre) $\sim$26\,d before and $\sim$1\,d after the detection of AT2017gfo.  The images are oriented with North up and East to the left and are cropped to 2~arcmin on a side, with the position of AT2017gfo  marked. The image taken on July 22, 2017 is in i-band, the image taken on August 18, 2017 (where the transient is visible) is in r-band. }
\label{figure: image SSS17a}
\end{figure}

Archival images at the coordinates of AT2017gfo were found from the SkyMapper Southern Sky Survey and the SkyMapper Transient Survey from August 8, 2015 to July 22, 2017.  We found no evidence of a pre-existing source or variability in the images coincident with AT2017gfo to a a 95\% upper limit of $i\sim$19.6 and $r\sim$20.5 \citep[Figure~\ref{figure: image SSS17a};][]{GCN21542}.

Imaging of the LVC skymap started at 2017-08-18 09:04:56 in the \textit{uvgriz} filters with $t_{\mathrm{exp}}$=100\,s. The images of AT2017gfo were taken between 2017-08-18 09:16:58 and 2017-08-18 10:00 UT in all bands (Figure~\ref{figure: lc short}).  The observations were taken at an airmass above 2 and roughly half of the primary mirror  was vignetted by the telescope dome.  As a result of dome seeing and high airmass, the images have a seeing FWHM of 3.5--6~arcsec in $i$/$z$-bands to $u$-band.  Nevertheless, the transient AT2017gfo was immediately confirmed visually on raw frames in all six bands.  Preliminary photometric AB magnitudes are given as $u=17.9\pm$0.15, $v=17.9\pm$0.10, $g=17.76\pm$0.05, $r=17.20\pm$0.05, and $i=16.0\pm$0.30, respectively \citep{GCN21560}.  

Observations of the source continued between 2017-08-18 to 2017-08-22, at which point AT2017gfo could no longer be visually identified in $uvg$ bands.  Imaging was attempted again between 2017-08-28 and 2017-09-03 to obtain images for host galaxy subtraction, but was unsuccessful.  A total of 83 successful exposures were taken with exposure times of 100\,s for bands \textit{griz} and up to 300\,s, for \textit{uv}.  Host galaxy images in all filters are planned  when the target re-appears from behind the Sun. 

\subsubsection{AST3-2}
The Antarctic Schmidt Telescope (AST3) project comprises three 68\,cm (50\,cm non-vignetted aperture) equatorial-mount telescopes located at the Kunlun Station at Dome A, Antarctica \citep{Cui2008}.

The second of the AST3 telescopes, AST3-2, employs a 10K$\times$10K STA1600FT camera with a pixel scale of 1 arcsec/pixel and a 4.14 $\deg$ field of view.  The AST3-2 observations presented in this paper were performed as part of the DWF program (PI Cooke).  Most facilities following AT2017gfo were only able to monitor the source for 1--2\,hr per night as a result of its position near the Sun.  The location of AST3-2 is advantageous in that it can monitor the source over longer periods of time as the source moved low along the horizon.  The disadvantages are that the source was always at high airmass and the dark Antarctic winter was ending.

Observations targeting the GW counterpart AT2017gfo span from 2017-08-18 to 2017-08-28 in SDSS-$i$ filter.  A total of 262 exposures were acquired, each with an exposure time of 300\,s per image, except for the initial 5 images having exposure time of 60\,s, with approximately 54\,s between exposures.  AST3-2 detected AT2017gfo on 2017-08-18 with an average $i$-band magnitude of $17.23^{+0.22}_{-0.21}$, $17.61^{+0.16}_{-0.16}$, and $17.72^{+0.18}_{-0.17}$ from coadded images.  The uncertainties of these measurements include the 0.088\,mag errors of the zero-point calibration.  The AST3-2 circular, \cite{GCN21883}, reports $g$-band magnitudes, however, this must be corrected to the $i$-band magnitudes that we report here.  Detections and upper limits estimated in the following observations are presented in Figure\,\ref{figure: lc short} and Table\,\ref{table: obs and photometry AST3}.

\subsubsection{Zadko}

The 1--m Zadko Telescope \citep{Coward2010} is located just north of Perth in Western Australia.  The CCD imager has a pixel scale of 0.69 arcsec/pixel (binning $1\times1$) resulting in a field of view of 0.15\,deg$^2$ and reaches an approximate limiting magnitude of 21 in the R-band in 180\,s.

The TAROT - Zadko - Aures - C2PU collaboration (TZAC) joins the efforts of partners located in Australia (Zadko), France (with TAROT telescopes in France, Chile and La R\'eunion Island, C2PU in France), and Algeria (Aur\`es Observatory, under construction). The initial position of GW 170817 was monitored using the TCH (TAROT-Chile) 25cm rapid robotic telescope prior to Zadko imaging.  

Zadko observations of AT2017gfo commenced on 2017-08-19 10:57 and extended until 2017-08-26 11:43 in the Clear (C) and r filters, with 120\,s exposures and $2\times2$ binning.  The object was observed for $\sim$1\,hr at the onset of dusk each night, until its low elevation precluded observations. 

We stacked all images taken each night to increase the signal to noise ratio under the assumption that the brightness of the object does not vary significantly during one hour.  As AT2017gfo is located at 10\,arcsec from the nucleus of NGC 4993 (i.e. 7\,pixels), the background varies steeply.  For accurate photometry, a galaxy reference image without AT2017gfo was subtracted to retrieve a flat background.  The reference image was created from the stack of images taken on the last night (i.e., 9 nights after the the GW trigger) when the source was no longer visible.  The photometry was performed on the subtracted image taking the point spread function (PSF) of the star NOMAD-1 0666-0296321 (RA=197$^\mathrm{h}$28$^\mathrm{m}$44.96$^\mathrm{s}$, Dec=--23$^\circ$21$^\prime$49.70$^{\prime\prime}$ J2000.0, $m_R$=15.580).  Photometric results are presented in \cite{GCN21744}, Figure\,\ref{figure: lc short}, and Table\,\ref{table: obs and photometry Zadko}.  


\subsubsection{University of Virgin Islands Etelman Observatory}
The Virgin Islands Robotic Telescope (VIRT) is a 0.5\,m Cassegrain telescope located the the Etelman Observatory in the U.\,S.\,Virgin Islands.  The observations with VIRT presented in this paper were performed in association with the DWF program.  VIRT is equipped with a Marconi 42-20 CCD imager that has a pixel scale of 0.5 arcsec/pixel, a field of view of 0.11\,deg$^2$, and imaging in the UBVRI and ND filters.

\begin{figure*}
\begin{center}
\includegraphics[width=2\columnwidth]{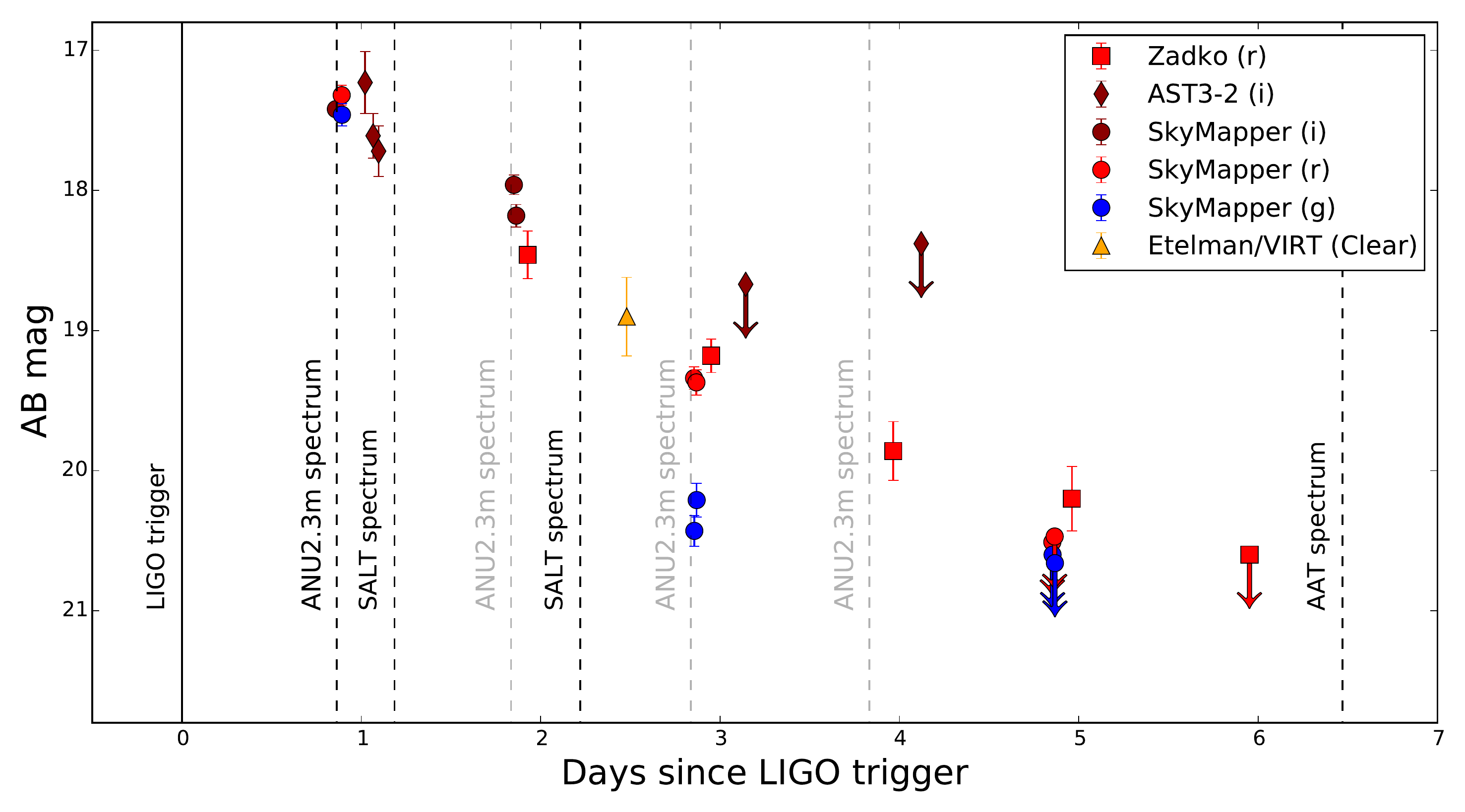}
\caption{Optical light curve of AT2017gfo for the first week after the GW detection obtained with the AST3-2, SkyMapper (SM), Zadko, and Etelman/VIRT telescopes.  Down-arrows indicate upper limits.  Note that the evolution at bluer bands is faster than the evolution at redder bands.  Dashed vertical lines indicate epochs when spectroscopy was acquired.  Spectra analysed in this work and presented in Figure~\ref{figure: spectra} and Figure~\ref{figure: host} are indicated in black, whereas spectra marked in grey are to be analysed at a later time. }
\label{figure: lc short}
\end{center}
\end{figure*}

Observations of AT2017gfo commenced on 2017-08-19 23:19 in the R and Clear (C) filters.  At approximately 2017-08-19 23:54, a potential counterpart was observed in the C filter.  Calculation of the precise source magnitude is limited due to the galaxy contamination in the observing band \citep{GCN21609}.  Additional observations were carried out on 2017-08-20 00:12 and 2017-08-22 00:00 with the C filter, where a possible first detection of the source was made on 2017-08-20 m$_C$ = 18.90 $\pm$0.28 (Fig.~\ref{figure: lc short}).  Inclement tropical weather (hurricane Irma, followed by hurricane Maria) delayed full analysis of the observations, however, the measurements made to date are listed in Table\,\ref{table: obs and photometry VIRT}.


\subsubsection{The Desert Fireball Network}
The Desert Fireball Network \citep[DFN,][]{Day2016} is a network of 50 remote cameras located in the Western and South Australian desert designed for the detection and triangulation of Fireballs and bright meteors.  Each DFN camera consists of a Nikon D800E camera equipped with a Samyang 8mm f/3.5 UMC Fish-eye CS II lens.  The cameras capture full sky images with a cadence of 30\,s from sunset to sunrise every night of the year.

Observations from Wooleen Station are available from two minutes before the GW170817 trigger and, as as result, DFN is the only optical facility imaging the source during the GW detection.  
Between 12:39:28 and 12:49:28, the host galaxy was observed at an elevation of 20\,deg. Initial analysis of the images finds no persistent or transient sources in a 3\,deg radius of NGC~4993, to a limiting magnitude of mag$_{v}$=4 \citep{GCN21894}.  Further calibration and analysis have brought this limiting magnitude down to mag$_{v}$=6.

\subsubsection{ESO VLT/NACO mid-IR}
The ESO Very Large Telescope (VLT) consists of four 8.2-m telescopes located at the Paranal Observatory in Chile.  Observations were made with the NACO instrument \citep{Lenzen2003,Rousset2003} on the VLT UT1 Antu telescope.  The system allows for adaptive-optics and natural seeing imaging over J, H, Ks, L$^\prime$, and M$^\prime$ filters, as well as providing Wollaston polarimetry and coronography in L$^\prime$.  The 5$\sigma$ limiting magnitudes are given as J=24.05, H=24.05, Ks=23.35, L$^\prime$=18.55, and M$^\prime$=15.15 in 1\,hr.  These observations were initially proposed as Director's Discretionary Time (PI Cooke, Baade) as part of the DWF program to be made immediately available to the LVC community.  However, the observations were finalised and executed by ESO, and made available to the LVC community.

Observations in the $L^\prime$-band (3.8\,$\upmu$m) were attempted on each night between 2017-08-24 and 2017-09-04.  Due to the proximity to the Sun and scheduling constraints, the target was observed during twilight (at UT 22:45--23:20) at airmass 1.5--1.6.  Weather and inaccurate pointing during the first nights resulted in the data from the four nights of August 25, 26, 27, and September 1 being analysed.  A pixel scale of 27\,mas\,pix$^{-1}$ was used for a total field of view of 27$\times$27 arcsec.  Observations were made in natural seeing mode with integration times 126$\times$0.2\,s per jitter point (with a 3 arcsec throw per axis), for a total of 15\,min, 19\,min, 14\,min, and 11\,min per night.  HD\,205772 was observed as a flux standard on August 28.  The data were reduced by a custom script in a standard way, correcting for sky variance by combining the jittered observations and de-striping by median filtering each detector quadrant separately.

\begin{figure}
\begin{center}
\includegraphics[width=\linewidth]{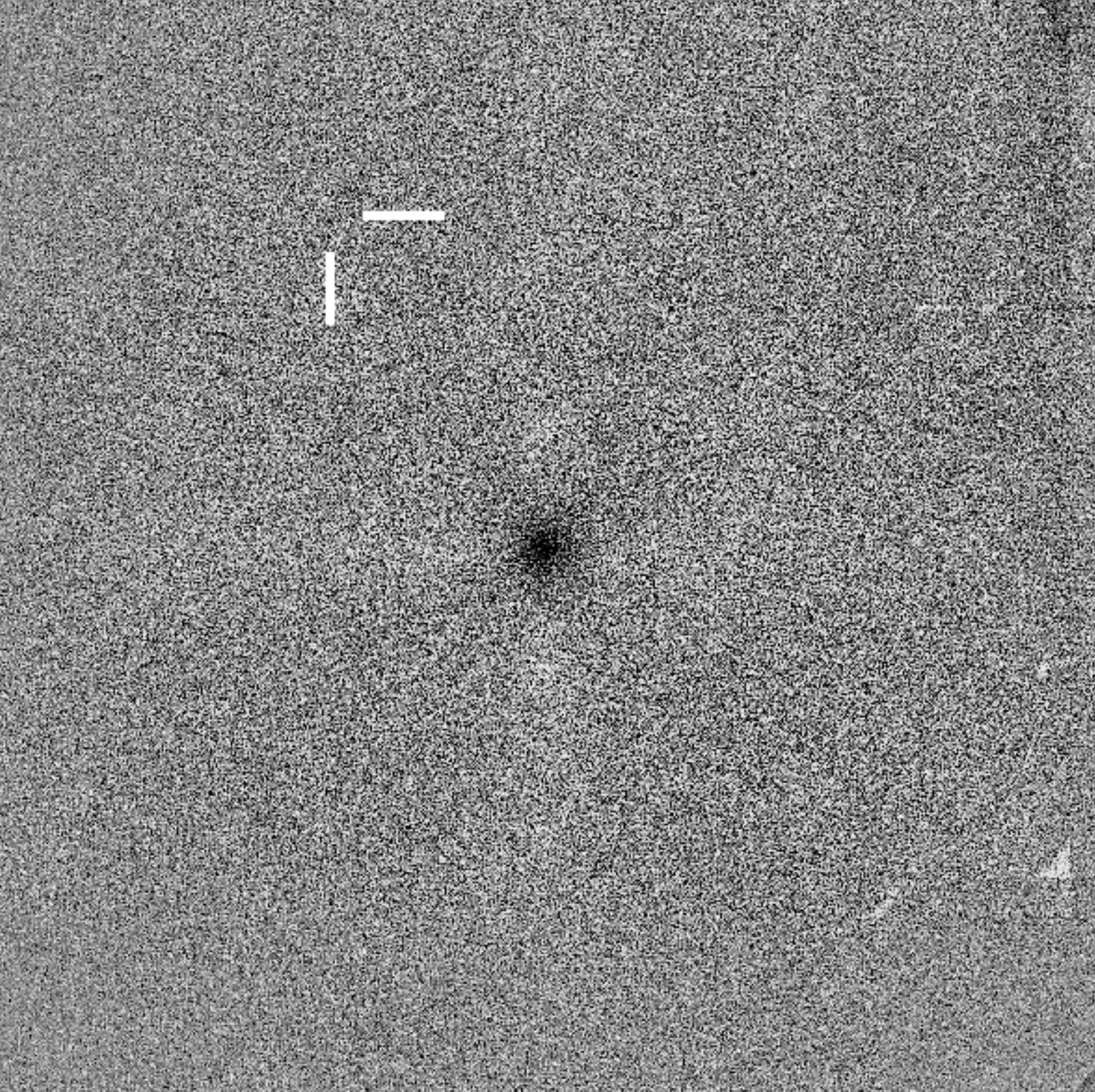}
\caption{Stacked NACO image of NGC~4993 (27$\times$27 arcsec), with the location of AT2017gfo marked. The image is oriented with North up and East to the left. The image is the combination of observations taken over four nights and no significant source was found to the detection limits of $L^\prime$ = 15.3, 5$\sigma$.}
\label{figure: NACO image}
\end{center}
\end{figure}

No sources apart from the NGC\,4993 nucleus were detected in the field (Figure~\ref{figure: NACO image}).  The detection limits were estimated from the background noise assuming a conservative point-spread function corresponding to the detected galactic nucleus at approximately 0.5 arcsec FWHM, using a circular aperture of 1 arcsec (40 pix) radius.  For the nights of August 25, 26, 27, and September 1, the 5$\sigma$ detection limits in $L^\prime$ are 14.5, 14.8, 14.5, and 14.3\,mag, respectively, with a combined limit of 15.3\,mag. 

\subsubsection{ESO VLT/VISIR mid-IR}
Imaging observations in the mid-IR were also made with the VISIR instrument \citep{Lagage2004} on the ESO VLT UT3 Melipal telescope.  Similar to NACO above, the observations were executed by ESO and made available to the LVC community.  VISIR provides an imaging field of view of $38^{\prime \prime} \times 38^{\prime \prime}$ with a plate scale of $0.045^{\prime \prime}$ per pixel.  AT2017gfo was observed on 2017 August 23, 2017 August 31, September 1, 2017, and 2017 September 6, 2017 with the J8.9 filter (central wavelength $8.72\,\upmu$m).  Total on-source integration times were 44.8, 17.5, 12.2, and 44.8 minutes, respectively.  Chopping and nodding in perpendicular directions with $8^{\prime \prime}$ amplitudes were used to remove the sky and telescope thermal background.  No source was detected to a limiting mag of J8.9 $\sim$7--8 (Table~\ref{table: obs and photometry VLT}).  Details of the observations can be found in \cite{Kasliwal2017}.

\subsection{Optical/Near-Infrared Spectroscopy}
Observations of AT2017gfo and the galaxy NGC 4993 were taken in the optical via longslit, fibre, and integral field unit (IFU) spectroscopic modes.  Both Australian and Australian partner observational programs participated in the spectroscopic follow up of AT2017gfo.  Details of the instruments and observations are provided below.  

\subsubsection{ANU2.3/WiFeS}
The Australian National University (ANU) 2.3-m telescope is located at Siding Spring Observatory in New South Wales, Australia.  It includes the dual-beam, image-slicing, integral-field echelle spectrograph \citep[WiFeS,][]{Dopita2007} which can simultaneously observe spectra over a 25$^{\prime\prime}\times$38$^{\prime\prime}$ field of view.  WiFeS has a spectral range extending from 3300 to 9800\,\AA, which can be observed either in a single exposure with a resolution of R = 3000, or in two exposures with R = 7000, depending on the choice of low- or high-resolution grating configurations, respectively.  The observations were done using Director's Discretionary Time.

\begin{figure}
\includegraphics[width=\linewidth]{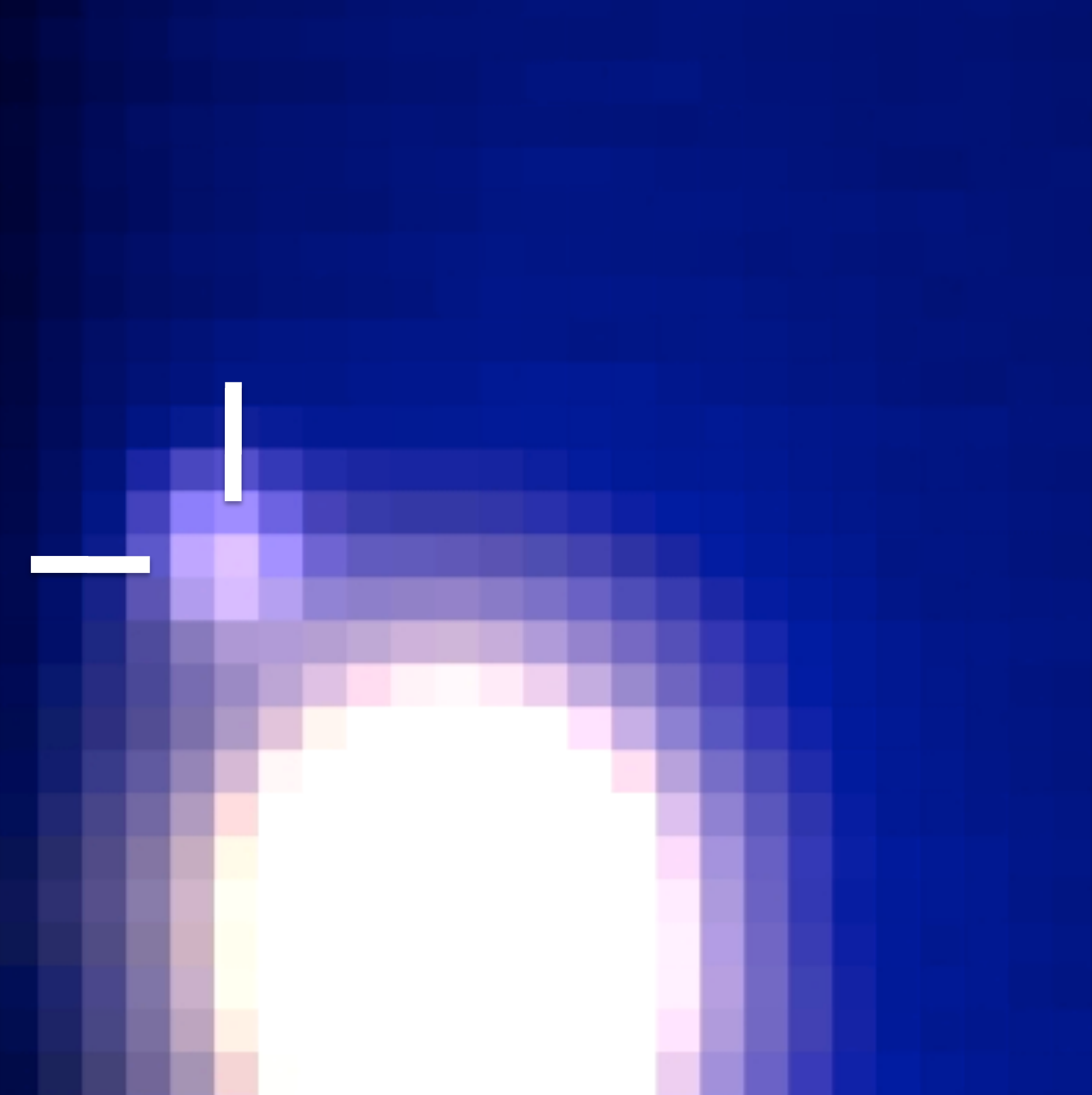}
\caption{WiFeS IFU collapsed data cube image (cropped to $\sim$25$\times$25 arcsec) of NGC~4993 and AT2017gfo (marked).  The image combines the data from both beams taken on 2017-08-18. The transient is noticeably bluer than the host galaxy.}
\label{figure: WiFeS image}
\end{figure}

Spectroscopic observations began on 2017-08-18 at 09:24:25 and 09:40:25 with a wavelength range of 3200--9800\,\AA.  Each observation had an exposure time of 15\,min.  The reduced spectrum shows a blue, featureless continuum peaking near 4500 \AA\ (Figure~\ref{figure: spectra}).  The observations continued for two further nights with the same configuration but a larger number of exposures to increase signal for the fading source.  The last exposures were taken on 2017-08-21 at times 08:40:58, 09:13, and 09:29 with a wavelength range of 3200--7060\,\AA, again with exposure times of 15\,min.  A WiFeS collapsed data cube image is shown in Figure~\ref{figure: WiFeS image}.

\subsubsection{SALT/RSS}
Opitical spectroscopsy of AT2017gfo was obtained using the Robert Stobie Spectrograph \citep[RSS,][]{Burgh2003} on the 10\,m-class Southern African Large Telescope (SALT) located in Sutherland, South Africa.  The observations were taken with Director's Discretionary Time initiated as part of the DWF program.  The RSS  is a spectrograph covering the range 3200--9000\,\AA\ with spectroscopic resolutions of R = 500-–10000.  The observations were performed using the PG0300 grating at an angle of 5.75\,$\deg$ and the 2\,arcsec slit.  Data taken on  2017-08-18 at 17:07 and 2017-08-19 at 16:59 \citep{GCN21610} had exposure times of 433\,s and and 716\,s, respectively.  Due to the visibility limitations of SALT, the data were acquired in early twilight and are heavily contaminated with a high sky background.  Spectral flux calibration standards were also observed on the same night.  

\begin{figure*}
\centering
\includegraphics[width=2\columnwidth]{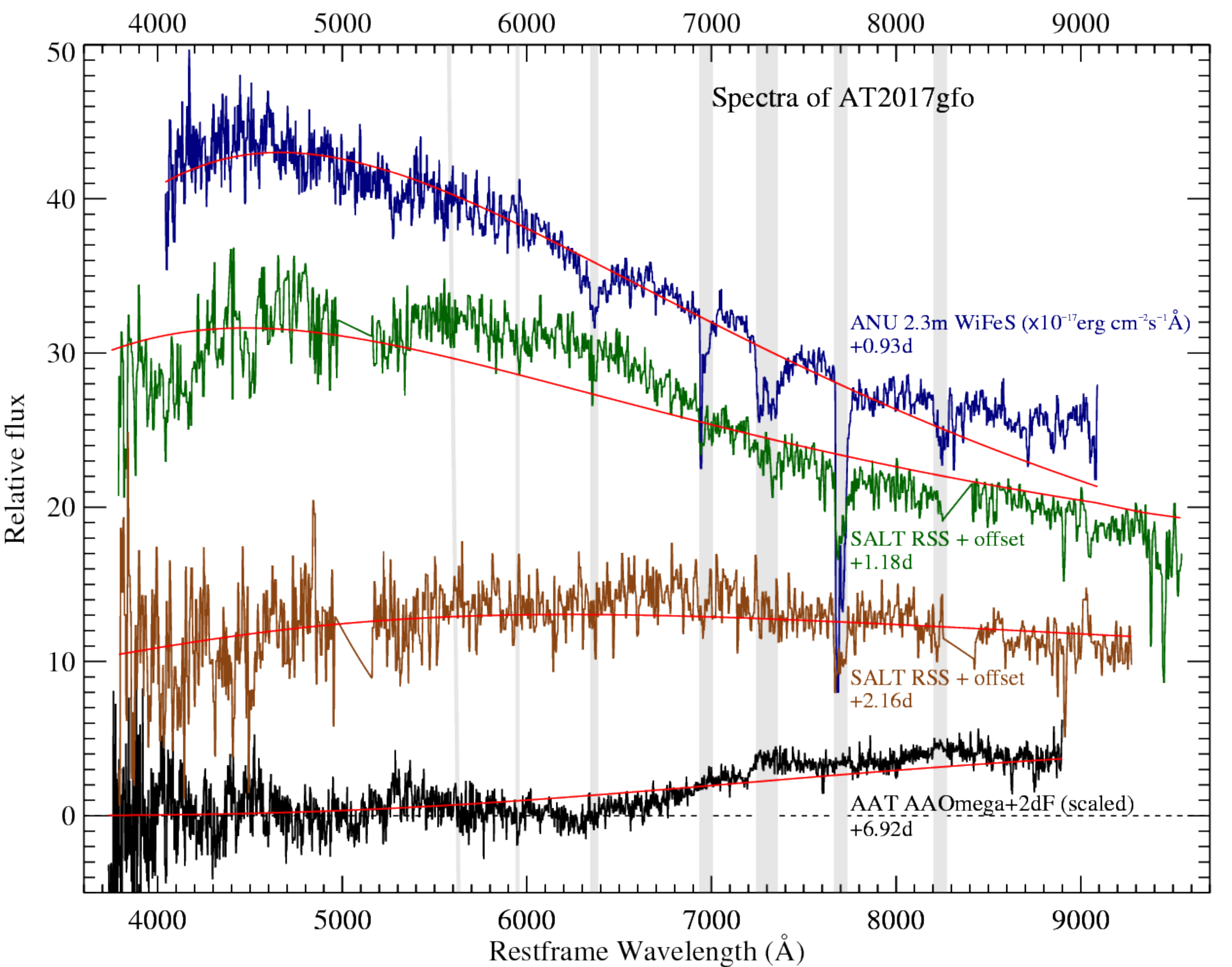}
\caption{The rapid spectral evolution of AT2017gfo.  The ANU 2.3m WiFeS, SALT RSS (2 spectra), and AAT AAOmega+2dF spectra obtained at 0.93\,d, 1.18\,d, 2.16\,d, and 6.92\,d, respectively, after GW detection are shown and labelled.  Vertical grey bands denote telluric features that are not well removed in some spectra.  Blackbody model fits (red curves) over the full spectra result in temperatures of 6275~K (WiFeS), 6475~K and 4700~K (RSS), and 2080~K (AAOmega).  Peaks in the WiFeS, RSS, and AAOmega continua correspond to $\sim$6400~K, $\sim$5600~K, $\sim$4400~K, and $<$ 3200~K, respectively.}
\label{figure: spectra}
\end{figure*}

Basic CCD reductions, cosmic ray cleaning, wavelength calibration, and relative flux calibration were carried out with the PySALT package \citep{Crawford2010}.  Because of the changing pupil during SALT observations, only a relative flux calibration can be achieved.  In order to de-blend the sources, the flux from the host galaxy, the atmospheric sky lines, and the GW source were fit simultaneously using the astropy.modeling package \citep{Astropy2013}.  The reduced spectra appears to have a relatively blue, featureless continua as seen in Figure~\ref{figure: spectra}.  The data are also presented and interpreted in \cite{McCully2017} and \cite{Buckley2017}.

\subsubsection{AAT/2dF+AAOmega}\label{AAOmega}
The Anglo-Australian Telescope (AAT) is a 3.9\,m Equatorial-mount optical telescope located in New South Wales, Australia.  AAOmega is a dual-beam optical fiber spectrograph with 3700 to 8800\,\AA\ wavelength coverage and a spectroscopic resolution of R = 1700 \citep{Smith2004}.  We used AAOmega combined with the Two Degree Field (2dF) multi-object system which allows for simultaneous spectroscopic observations of up to 392 objects within a 2\,deg diameter field of view.  The observations were done as part of the DWF program and granted via Director's Discretionary Time while activating the newly commissioned AAT 2dF Target Of Opportunity (ToO) mode.  Fully configuring all 392 fibres takes $\sim$40 minutes and is too long for rapid follow-up of short-lived transient phenomena.  In rapid ToO mode, the 2dF software determines, from an existing fibre configuration, which fibres need to move to place a single fibre on the target and one on a guide star.  This capability enables configuration and observation within a few minutes and, in the case of AT2017gfo, 5 minutes between ToO activation and the commencement of the observations. 

AT2017gfo observations began on 2017-08-24 at 08:55:07 to 09:41:28 with exposure times of 600\,s each.  The data were processed using the OzDES pipeline \citep{Childress2017}.  Four exposures were analyzed, revealing an E/S0-like galaxy spectrum (Figure~\ref{figure: host}) with a weak red flux enhancement \citep{GCN21677}.  The source was isolated by subtracting the host galaxy using the SALT host galaxy spectrum \citep{McCully2017} extracted from the region of the galaxy near the source.  The SALT spectrum was cleaned over chip gaps and telluric line regions using the average value on either edge of each feature.  Finally, the SALT host and AAT AAOmega host+event spectra were scaled and subtracted (Figure ~\ref{figure: spectra}).  Subtracting two spectra with relative flux calibrations introduces uncertainties in the scalar offset.  Such subtractions do not significantly affect the form of the residual spectrum, but can provide a small affect on blackbody model fit results.  Although care was taken in the subtraction process, the two spectra introduce possible flux calibration differences from the different instruments and extraction techniques.  As a result, we stress that the spectrum presented here is meant to be indicative of the behaviour and temperature of the event at 6.92\,d, and suffers from the above caveats.  A proper host galaxy subtraction with the AAT AAOmega+2dF is planned when NGC 4993 becomes visible.

\subsection{Radio}
Five Australian and international radio facilities participated in this follow up campaign.  In this section we describe the role of each radio observatory that performed the follow up of GW170817 and/or AT2017gfo under Australian-led observing programs. 

\subsubsection{ATCA}
The Australia Telescope Compact Array (ATCA) is located at the Paul Wild Observatory in New South Wales, Australia.  It is an array of six 22\,m radio antennas, which can be configured with antenna spacings up to 6\,km. The array can observe in one of 5 observing bands spread between 1.1 and 105\,GHz.

We carried out ATCA observations on August 18, 21, 28, and September 5, 2017 under a target of opportunity program (CX391; PI: T. Murphy).  During the August observations, we targeted 53 galaxies identified to be located within the 90$\%$ containment volume of GW170817 \citep{GCN21537,GCN21559}.  The September 5 observation targeted only the optical counterpart, AT2017gfo and its host galaxy NGC 4993.  Table\,\ref{table: obs and photometry ATCA} presents a summary of the observations.

The August observations used two 2~GHz frequency bands with central frequencies of 8.5 and 10.5\,GHz and observed NGC 4993 using two frequency bands centered on 16 and 21\,GHz on August 18,  For the September observations we centered these two frequency bands on 5.5 and 9.0\,GHz.  The configuration of the ATCA changed over the course of the observations, with ATCA in the EW352 configuration for the August 18 observation and in the 1.5A configuration for all other observations.  

For all epochs and all frequencies, the flux scale was determined using the ATCA primary calibrator PKS B1934-638.  The bandpass response at 8.5 and 10.5\,GHz was determined using PKS B1934-638 and observations of QSO B1245-197 were used to calibrate the complex gains.  We used QSO B1921-293 to solve for the bandpass at 16.7 and 21.2\,GHz and observations of QSO B1256-220 were used to solve for the complex gains at these frequencies.  All of the visibility data were reduced using the standard routines in the \textsc{miriad} environment \citep{Sault1995}.

We used the \textsc{miriad} tasks \textsc{invert}, \textsc{clean}, and \textsc{restor} to invert and clean the calibrated visibility data from the August observations of the 53 targeted galaxies. We fit a single Gaussian to each of the 53 galaxies detected in our August observing epochs (Lynch et al. LVC GCN 21628, Lynch et al. LVC GCN 21629).  Comparing these observations, we find no transient emission above a 3$\sigma$ limit between 36 - 640\,$\mu$Jy. The measured flux densities for host galaxy NGC 4993 are listed in Table \ref{tab:ATCANGC}. The results from our observations of AT2017gfo are described in \citet{Hallinan2017}, including a detection on September 5 at 7.25\,GHz, with measured flux density of 25$\pm$6\,$\mu$Jy \citep{GCN21842}.


\subsubsection{ASKAP}

The Australian Square Kilometre Array Pathfinder (ASKAP) is a system of thirty six  12\,m phased-array feed receiver radio telescopes located in Western Australia.  The instrument covers a frequency range of 0.7 to 1.8\,GHz with a bandwidth of 300\,MHz. The field of view of is 30\,deg$^2$ at 1.4\,GHz, with a resolution of $\sim$30\,arcsec.

ASKAP performed imaging observations on 2017-08-19 05:34:32 (LVC GCN 21513) with 12 of the 36 antennas\footnote{As a result of ongoing commissioning}.  The 90\% LVC contour region (The LIGO Scientific Collaboration and the Virgo Collaboration 2017d) was covered with 3 pointings using an automated algorithm \citep{Dobie2017inprep} observed over the following 4 days.  We place an upper limit of $\sim$1 mJy on emission from AT2017gfo and its host galaxy NGC4993.

At the time of publication, 14 further single-beam observations of the AT2017gfo location were carried out with varying numbers of beams and antennas at different frequencies and bandwidths (subject to commissioning constraints).  These observations are undergoing processing, while further observations are ongoing.

ASKAP also searched the 90\% LVC uncertainty region at high time resolution for fast radio bursts \citep[FRBs][]{Lorimer2007} using the search algorithms described in Sec.\,2.3.5 to cover a dispersion measure range of 0--2000 pc cm$^{-3}$.  The observations were in `fly's-eye' mode with 7 antennas at a central frequency of 1320\,MHz \citep{Bannister2017}.  Observation times were 2017-08-18 04:05, 2017-08-18 08:57, and 2017-08-19 02:08, for a total duration of 3.6, 4.1, and 11.0\,hrs, respectively.  Above a flux density threshold of $\sim$40\,Jy/$\sqrt{w}$, there were no FRB detections (GCN21671), where $w$ is the observed width of the FRB in milliseconds.

\subsubsection{MWA}
The Murchison Widefield Array (MWA) is a system of 2048 dual-polarization dipole antennas organized into 128 tiles of 4$\times$4 antennas located in Western Australia.   MWA operates between 80 and 300 MHz \citep{Tingay2013} and has a resolution of several arcmin.  Operations with the original array (baselines up to 3\,km) with a compact configuration with maximum redundancy ceased in 2016.  The reduced baseline was used until mid-2017 at which point tiles with extended baselines up to 5\,km were installed for MWA Phase~II.

The telescope responded automatically to the LVC GCN \citep{Kaplan2015} but the initial LVC notice only included information from a single detector, so the telescope pointing was not useful.  Later we manually pointed the telescope and began observations on 2017-08-18 at 07:07 with only 40 tiles in a hybrid array with elements of the maximally redundant array and the original array.  Observations occurred daily from 2017-08-18 to 2017-08-22 with $75\times2\,$min exposures and then continued weekly.  The observations cover a 400\,deg$^2$ field of view at a central frequency of 185\,MHz and a bandwidth of 30\,MHz \citep{GCN21637}.  We see no emission at the position of NGC 4993 
with a flux density limit of 51\,mJy/beam (3$\sigma$ confidence) from the data taken on August 18, 2017 \citep{GCN21927}.  Later observations with more functioning tiles and longer baselines should have considerably improved performance.  \cite{Kaplan2016} discuss in detail the strategies to use MWA for finding prompt radio counterparts to GW events. 


\subsubsection{VLBA}
The Very Long Baseline Array (VLBA) is a radio interferometer consisting of ten 25\,m radio telescopes spread across the United States, and is capable of observing in one of 10 bands at frequencies between 1.2 and 96\,GHz.

The counterpart AT2017gfo and its host galaxy NGC 4993 were observed on three occasions under the Director's Discretionary Time project BD218, each with 6.5 hour duration.  The observations were performed from 2017-08-18 19:58 to 2017-08-19 01:34, 2017-08-20 18:31 to 2017-08-21 01:13, and 017-08-21 18:26 to 2017-08-22 01:09.  The central observing frequency was 8.7 GHz, with a bandwidth of 256 MHz and dual polarisation.  The source VCS1 J1258-2219, with a position uncertainty of 0.2 mas, was used as a primary phase reference calibrator, with NVSS J131248-235046 as a secondary calibrator.  An observing failure rendered the first epoch unusable, but the second and third epochs provided good data.

No source was detected within 0.5~arcsec of the position of AT2017gfo, consistent with the findings of both the VLA and ATCA instruments \citep[e.g.,][]{GCN21559, GCN21574, GCN21670}.
However, we are able to provide 5.5$\sigma$ upper limits of 125\,$\upmu$Jy/beam and 120\,$\upmu$Jy/beam at August 20, 2017 21:36 and August 21, 2017 21:36, respectively, while stacking the two images produces an upper limit of 88\,$\upmu$Jy/beam \citep{GCN21588, GCN21850}.

Imaging the core region of NGC 4993 identifies a sub-mJy radio source at the centre with coordinates RA=13$^\mathrm{h}$09$^\mathrm{m}$47.69398$^\mathrm{s}$ Dec=--23$^\circ$23$^\prime$02.3195$^{\prime\prime}$ (J2000).  The detection is consistent with either an unresolved source or a marginally-resolved source on a scale smaller than the VLBA synthesized beam (2.5$\times$1.0 mas).  The systematic uncertainties of our position are $\leq$ 1\,mas in both RA and DEC.  We find a 9$\sigma$ flux density of 0.22 mJy, and the {\em a priori} amplitude calibration available to the VLBA is accurate to the 20\% level.  If we assume the synthesized beam size of 2.5$\times$1.0 mas to represent a conservative upper limit on the size of the source, we infer a lower limit for the brightness temperature of 1.6$\times$10$^6$\,K.  An initial interpretation suggests the recovered brightness temperature is consistent with an AGN \citep{GCN21897}.  Comparison of the flux densities estimated by ATCA and VLA \citep[see Table\,\ref{tab:ATCANGC} and][]{Hallinan2017} to the VLBA value indicates that a considerable amount ($\sim$50\%) of the total source flux is contained within this mas scale component.

\subsubsection{Parkes}
\label{subsubsec: Parkes}
The Parkes Radio Telescope (Parkes) is a 64-m telescope located in Parkes, New South Wales, Australia.  Parkes operated in FRB search mode with the Multibeam receiver \citep{Staveley-Smith1996} and the BPSR backend \citep{Keith2010}.  The usable bandwidth is 340~MHz, in the range of 1182--1582~MHz.  If the neutron star merger produced a massive ($>2$M$_\odot$) neutron star instead of a black hole, it would be expected to possess a spin period close to the break-up velocity of $\sim$1 ms and potentially a large magnetic field generated during its formation.  Such objects (millisecond magnetars) are a potential source of FRBs or possibly even repeating FRBs \citep{Spitler2016,Metzger2017}.  The FRB should be detectable at S/N $>$ 100 with Parkes at the distance of NGC~4993, if appropriately beamed and not hidden by the ejecta from the merger.

A dedicated search for FRBs \citep{Keane2017} with dispersion measures ranging from 0--2000 pc~cm$^{-3}$ associated with AT2017gfo was performed on 2017-08-18 at 06:49:31 and 08:50:36 with 2-hr and 1-hr integration times, respectively, and again on 2017-08-20 at 01:44:32 and 02:50:14 with 1-hr integration times \citep{GCN21899,GCN21928}.  No FRBs were detected with a 7$\sigma$ limiting flux density of 1.4 sqrt(w/0.064) Jy sqrt(ms),
where $w$ is the observed pulse width of the FRB in ms.




\section{ANALYSIS}
\label{sec:models}

The observations presented here identified the optical transient on multiple epochs for the first $\sim$7 days after the LIGO trigger.  In Figure~\ref{figure: lc short} we present the the multi-band photometric light curve of AT2017gfo, observed in $g$-band (SkyMapper), $r$-band (SkyMapper, Zadko, Etelman/VIRT), and $i$-band (AST3-2, SkyMapper).  The multi-band measurements indicate a decay faster in $g$-band than in the $r$- and $i$-bands.  We processed and analysed four optical spectra acquired with ANU2.3m/WiFeS, SALT/RSS, AAT/2dF+AAOmega.  The subtraction of the host galaxy allows the signature of the transient to be identified and the spectral evolution to be assessed (Figure\,\ref{figure: spectra}).  In this section, we review the spectral evolution of AT2017gfo, describe the properties of the host galaxy NGC~4993, and assess the photometric evolution of the event compared to sGRB and kilonova models.  


\subsection{Spectral evolution of AT2017gfo}
\label{subsec: BB fits}
The ANU 2.3m (WiFeS), SALT (RSS), and AAT (AAOmega+2dF) spectra reveal a rapid evolution of the transient over $\sim$7\,d while maintaining relatively featureless continua.  As a coarse measure of the evolving spectral energy distribution, we fit a blackbody model to the spectra (Figure~\ref{figure: spectra}).  Continuum blackbody temperatures were calculated by fitting the observed spectra using the python {\tt scipy} package implementation of the non-linear least-squares Levenberg-Marquadt algorithm.  Spectra are corrected to rest-frame and for Milky Way line-of-sight extinction using the \cite{CCM1989} prescription and adopting R$_V$ = 3.1 and $E(B-V)$ = 0.12 and based on the dust maps of Schlegel, Finkbeiner, \& Davis (1998).

The model fits result in a temperature evolution from $\sim$6400~K to $\sim$2100~K in $\sim$7\,d.  The WiFeS spectrum is reasonably well fit by a $\sim$6300~K blackbody, with the peak in the spectrum continuum corresponding to $\sim$6400~K.  The curvature of the SALT spectrum is not well fit by a blackbody model, with the model fit producing a temperature of $\sim$6500~K, whereas the peak in the spectrum roughly corresponds to $\sim$5600~K.  The second SALT spectrum, taken at +2.16\,d, is reasonably well fit, producing a blackbody model fit of $\sim$4700~K, while the continuum peak corresponds to roughly 4400~K.  By day $\sim$7 the source is quite faint and host galaxy subtraction is less reliable.  The AAOmega+2dF spectrum at +6.92\,d is best fit by a blackbody model at $\sim$2080~K, but has the caveats stated in Section \ref{AAOmega}. 

\subsection{The host galaxy}
\label{sec: host galaxy}

\begin{figure*}
\centering
\includegraphics[width=2\columnwidth]{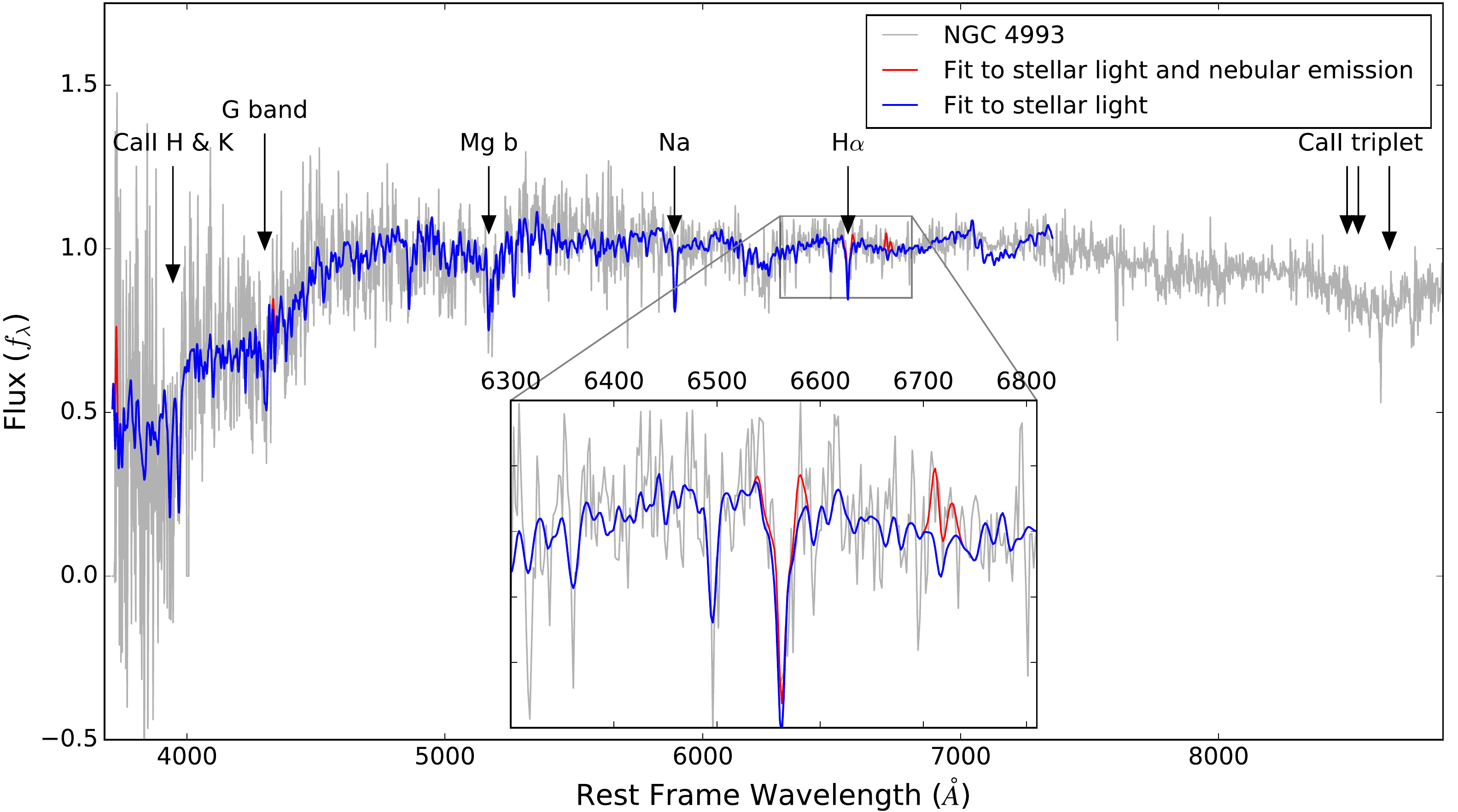}
\caption{AAT fibre spectrum of NGC 4993 in a 2~arcsec region at the position of AT2017gfo.  A fit to the stellar light (blue) and the stellar light and nebular emission (red) are shown.  The fits include the flux of AT2017gfo (at +6.92\,d) and the galaxy.  Several common atomic transitions are marked and a zoom-in of the H-$\alpha$ region is shown.  The spectrum is corrected for line-of-sight Milky Way extinction.}
\label{figure: host}
\end{figure*}

The AAT/2dF+AAOmega spectrum \citep[Figure~\ref{figure: host},][]{GCN21677} was acquired 6.92\,days after the LIGO trigger.  The fibre was centred on the transient position, but the spectrum is dominated by the light of the host galaxy.  Figure~\ref{figure: spectra} shows the galaxy-subtracted transient spectrum from the same observation. 

We use pPXF \citep{Cappellari2017} to fit the spectrum to 7300\AA\ (the extent of the MILES spectral template library) to estimate the metallicity, age, r-band mass-to-light ratio and velocity dispersion of stars in the region immediately surrounding AT2017gfo.  At the redshift of the host galaxy, the 2\,arcsec diameter of a 2dF fibre corresponds to a linear size of 400\,parsec. Assuming a spectral resolution of 4.5\,\AA\ (measured using night sky lines), the pPXF fit yields a velocity dispersion of 100\,km\,sec$^{-1}$, a stellar age of 10 billion years, a metallicity of [M/H] = $-0.2$, and an r-band mass-to-light ratio of 4.  Evidence for H$\alpha$ emission in the pPXF fit is very weak, measured at EW = $-0.2$\,\AA, but is consistent with zero.  The environment in the location of the transient is consistent with an old, passively evolving stellar population with no on-going star formation. 


The above assessment of NGC~4993 in the region of the source is consistent with the report of \citet{GCN21645} for the central 6\,arcsec region of the galaxy based on spectroscopy from the 6dFGS \citep{Jones2009}.  The central stellar velocity dispersion of 163 km/s \citep{Ogando2008} predicts a central black hole mass of $M_{\mathrm{BH}} = 10^{7.7}$\,M$_\odot$\ \citep{GCN21669}, which can be compared against estimates based on the radio properties of the central source.  The compact radio emission detected by VLBA in the central region of NGC~4993 with a brightness temperature exceeding $10^{6}$~K indicates the presence of a low luminosity active galactic nucleus (LLAGN), allowing us to estimate the black hole mass using the fundamental plane of black hole activity \citep[e.g.][]{Plotkin2012,Merloni2003,Falcke2004}.  The VLBA flux density was measured to be $0.22\pm0.04$~mJy at 8.7 GHz, which (assuming a flat spectral index) gives a 5 GHz radio luminosity of $\left( 2.1\pm0.04 \right) \times 10^{36}$ erg s$^{-1}$, while the X--ray luminosity as measured by SWIFT is $5.6^{+2.4}_{-1.9}\times10^{39}$ erg/s \citep{GCN21612}.  The radio spectral index is consistent with being flat or slightly negative (as can be seen from the ATCA results shown in Table~\ref{tab:ATCANGC}); the results are insensitive to small variations in this parameter.  Using the relationship described in \citet{Plotkin2012}, we obtain a predicted central black hole mass of $10^{7.8\pm0.3}$\,M$_\odot$, in good agreement with the velocity dispersion estimate.

\citep{GCN21645} also state that the nuclear dust lanes evident in the HST ACS images \citep{GCN21536,Pan2017} may be the product of a galaxy-galaxy merger that occurred as long as several Gyr ago.  We note that a wet galaxy merger (to produce the visible dust) implies that the binary progenitor of AT2017gfo might have originated in the merging galaxy and not necessarily in the main early-type host.  Such an origin could permit a shorter binary neutron star inspiral time than would be plausible for a massive galaxy with no recent star formation.  Previous sGRB hosts with possible kilonovae are often low-mass, blue star forming galaxies \citep{Tanvir2013,Fong&Berger2013}, though 20--40\% of sGRBs occur in early-type galaxies \citep{Fong2013}.  The diversity of possible host galaxies for neutron star merger events therefore needs to be kept in mind when searching for the counterparts of future GW events. 




\subsection{Comparison with GRB afterglow and kilonova models}

The optical data we acquired, alone, can give insight on the nature of the transient event.  First, we explore the GRB afterglow scenario in order to test the possibility that AT2017gfo behaves as a ``standard'' on-axis GRB in the optical, specifically using the \cite{Granot1999} and \cite{GranotSari2002} models.  Secondly, we investigate the kilonova scenario by comparing the data we acquired with three possible models \citep{Tanaka2013,Hotokezaka2013,Barnes2013,Metzger2015}.  In Figure\,\ref{fig:TanakaNSNS} we overlay the results we obtain to our data.

\subsubsection{GRB afterglow}

We investigate the GRB afterglow scenario using the Granot \& Sari \citep[][G02]{Granot1999,GranotSari2002} formulation for a relativistic blast wave in an ISM environment.  Far from the sites of the break frequencies of the GS02 spectra, each power law segment becomes asymptotic.  In particular, we can assume that the frequency of our optical observations, $\nu_{opt}$, relates to other characteristic frequencies as $\nu_{sa} < \nu_m < \nu_{opt} < \nu_c$, where $\nu_{sa}$ is the self-absorption frequency, $\nu_m$ is the minimal electron synchrotron (or peak) frequency, and $\nu_c$ is the frequency at which an electron cools over the dynamical time span of the system.  In this region of the spectrum we can approximate the spectral flux density as $F_\nu \propto t^\alpha$.  Simultaneous X--ray or radio measurements would help to constrain the locations of the break frequencies of the spectrum.

We calculate the index $\alpha$ by $\chi^2$ minimization of the Zadko telescope r-band data points and we find $\alpha=-1.73\pm0.10$, in addition we derive an electron power law index $p = 1+\frac{4}{3}\alpha$ (G02) to determine $p = 3.31\pm 0.13$.  This value is higher than historical sGRBS \citep[see][for a decadal review]{Fong2015}, where the median value of $p$ is found to be $\langle p \rangle = 2.43^{+0.36}_{-0.28}$.  In a classical sGRB scenario, our calculated $p$ could be interpreted as (i) emission is not a spherically isotropic blast wave \citep{Sari1999} giving a larger temporal decay slope than historical sGRBs \citep{Fong2015} or (ii) evidence that the jet itself may be structured \citep{Salafia2015}.

We use the isotropic gamma-ray energy measured with {\it Fermi} $\left\langle E_{\gamma,\mathrm{iso}}\right\rangle \approx (3.0\pm0.6) \times 10^{46}$\,erg \citep{Goldstein2017} to constrain our parameter space, assuming that E$_{\gamma,\mathrm{iso}}$ $\approx$ E$_{\mathrm{K},\mathrm{iso}}$ \citep{Frail2001}.  In this way, we find an unrealistically high circumburst number density $n \sim 10^{13}$\,cm$^{-3}$.  Placing these values back into the GS02 models produces the solid black lightcurve as shown in Figure~\ref{fig:TanakaNSNS}.  
The extremely high circumburst number density and the poor match with our observed optical data rule out the optical emission being the afterglow of a ``standard" on-axis sGRB.  This conclusion supports the idea that GRB~170817A was detected off-axis, as suggested by the {\it Fermi} detection of a sub-luminous GRB \citep{Goldstein2017} and the lack of any prompt X--ray afterglow detection \citep{GCN21572}.

\subsubsection{Kilonova models}

\begin{figure*}
\begin{center}
\includegraphics[width=1.2\columnwidth]{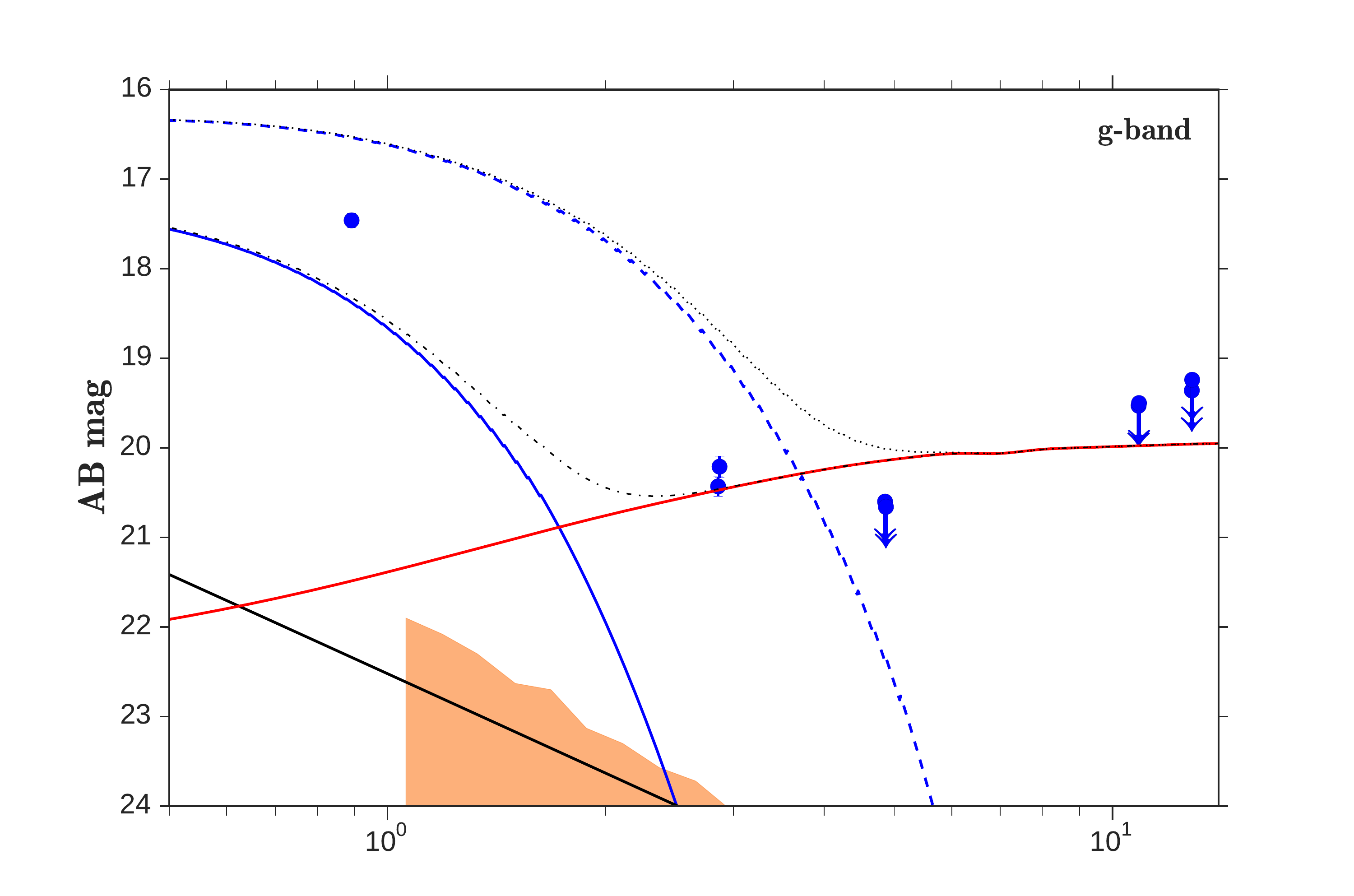}\vspace{-0.37cm}
\includegraphics[width=1.2\columnwidth]{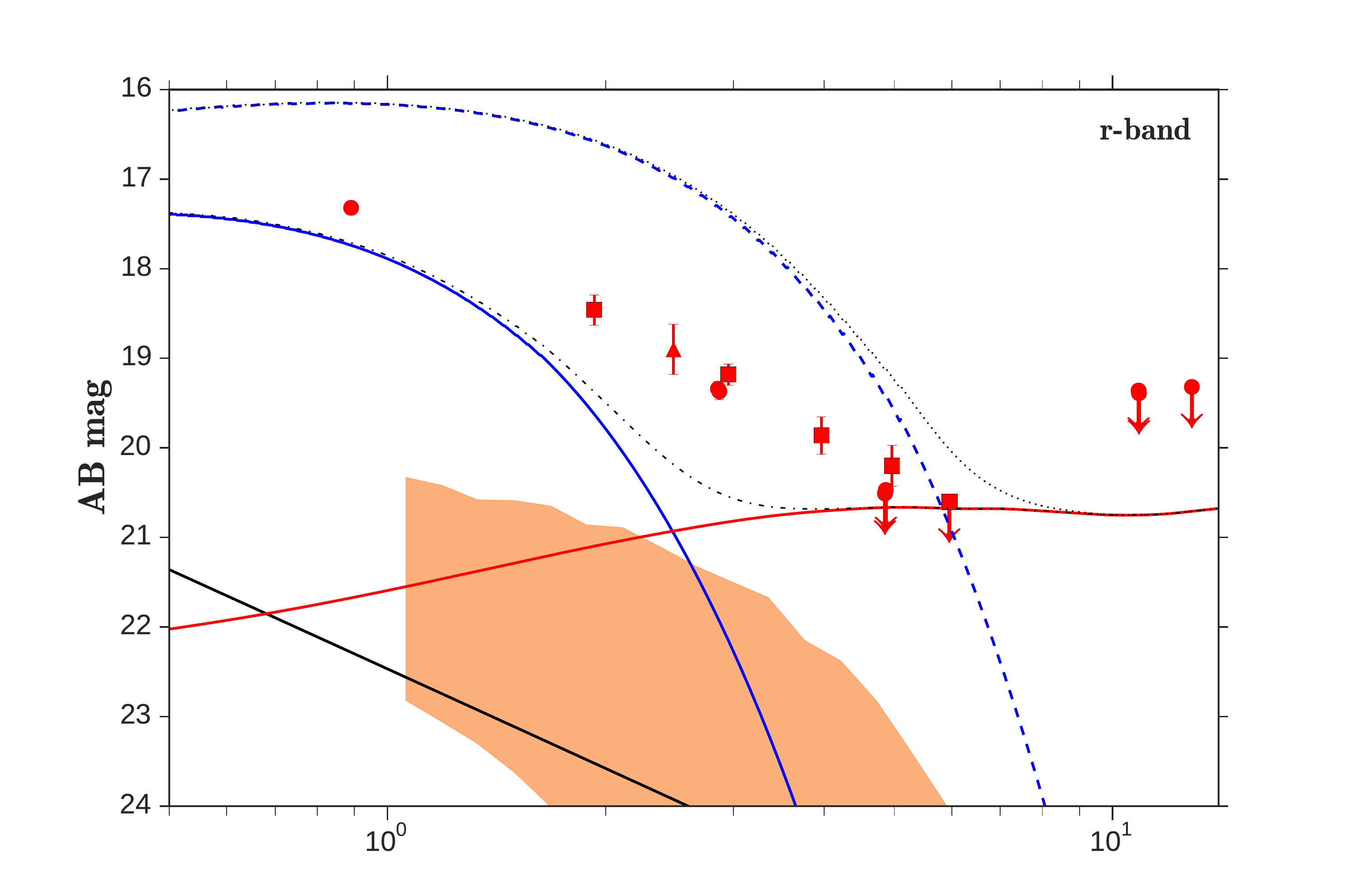}\vspace{-0.37cm}
\includegraphics[width=1.2\columnwidth]{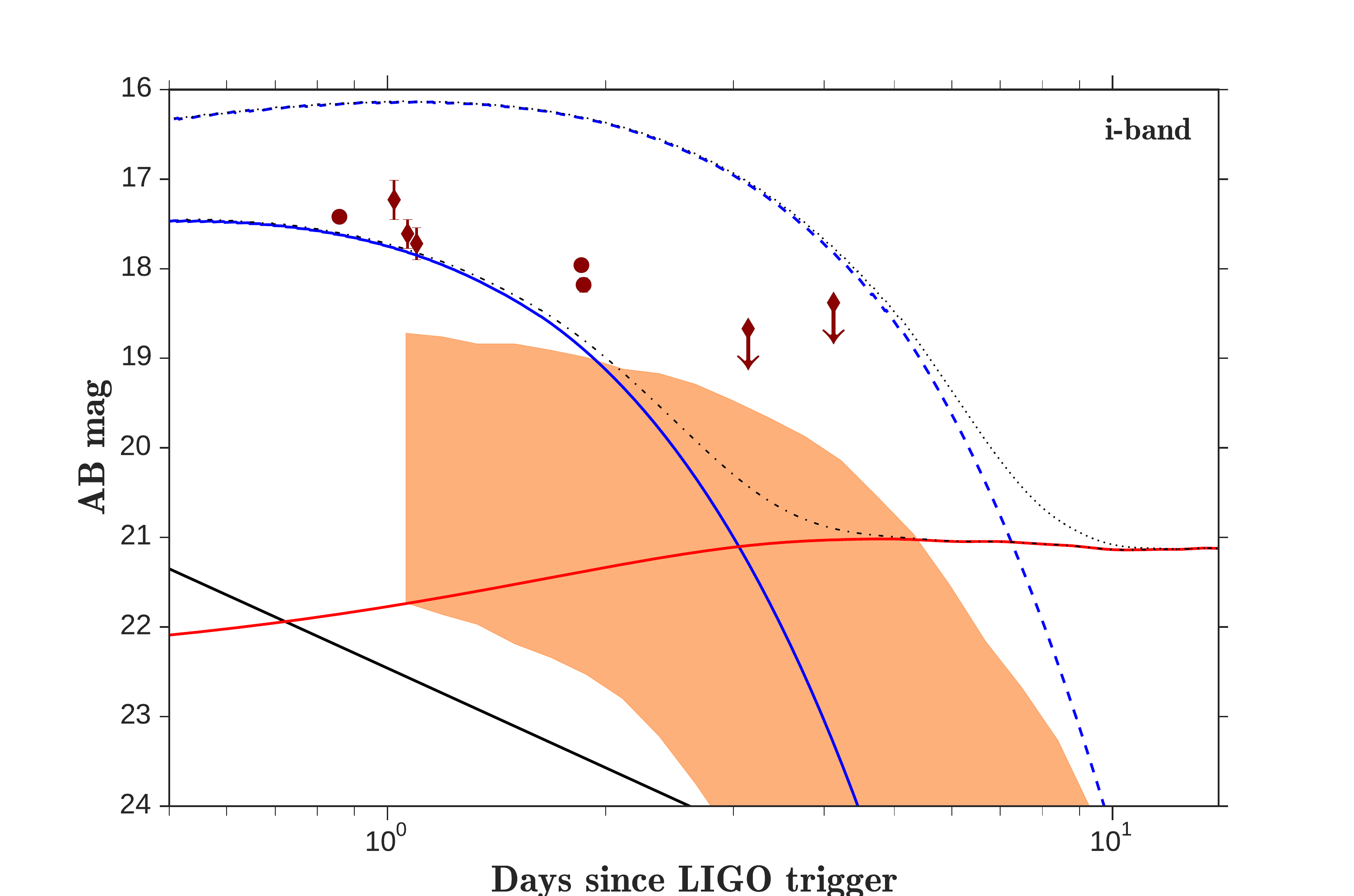}
\caption{Comparison of models to optical photometry with Zadko (squares, $r$-band), AST3-2 (diamonds, $i$-band), Etelman/VIRT (triangles, C-filter presented in the central panel), and SkyMapper (circles, {\it gri}-bands). The solid black line is the GS02 model of a short GRB afterglow.  The dark orange region represents the kilonova model by \cite{Tanaka2013}.  The red solid line represents the \cite{Barnes2013} model for $^{56}$Ni+r-process opacities.  The blue lines represent the free neutron-powered blue precursor \citep[solid: $v_{ej}=0.2c$, $M_{ej}=0.01\,M_{\odot}$; dashed: $v_{ej}=0.2c$, $M_{ej}=0.1\,M_{\odot}$][]{Metzger2015}, while the black dashed  and dot-dashed lines represent the \cite{Metzger2015} and \cite{Barnes2013} models together. The figure is organised in three panels, presenting photometry and overlaid models in $g$-band ({\bf top}), $r$-band ({\bf centre}), and $i$-band ({\bf bottom}). }
\label{fig:TanakaNSNS}
\end{center}
\end{figure*}

We compare our data with three standard models describing inherent kilonova emission.  In particular, we consider the case of r-processes in the ejecta from binary neutron stars mergers in the ``TH13" formulation \citep{Tanaka2013,Hotokezaka2013} for a range of NS equations of state, the \citep[``B\&K13" model,][]{Barnes2013}, and free neutron-powered blue precursor to the kilonova emission \citep[``M15",][]{Metzger2015}.  We plot the expected {\it gri}-bands light curves for all these models in Figure~\ref{fig:TanakaNSNS}. 

{\bf TH13 model:}  We calculate the expected light curves using the TH13 kilonova {\it gri}-bands light curves for a source located at $D_L$=40\,Mpc and for a variety of NS equations of state, specifically APR4-1215, H4-1215, Sly-135, APR4-1314, H4-1314.  We calculate the light curves for polar view angles, where the magnitudes are K-corrected in the rest frame using a standard $\Lambda$CDM cosmology with $H_0$=70 km/s/Mpc, $\Omega_m$=0.3, and $\Omega_\Lambda$=0.7 \citep{Hotokezaka2013}.  The results lie within the solid orange regions in Figure~\ref{fig:TanakaNSNS} and show a fainter emission than we observed.  The results are to be expected, as the spectra (Figure~\ref{figure: spectra}) are characteristic of a blue transient - at least in the first few days after the merger - while the TH13 model predicts a transient peaking at near-IR wavelengths.  The ``mismatch" between our measurements and the TH13 models reduces at late times and at redder bands (from $g$ to $i$), but only a longer monitoring of the source could indicate whether the transient can be dominated by r-processes at late times.

{\bf BK13 model:} In the BK13 model, the ejecta have an opacity similar to r-process material, made up of heavier lanthanide-group elements generated from dynamical ejection, and material made up of $^{56}$Ni which is ejected via disk winds.  These cases predict an emission peaking in the near-IR and optical, respectively \citep{Barnes2013}.  We show the results for the emission expected from $^{56}$Ni in Figure~\ref{fig:TanakaNSNS} as a dashed grey line.  At late times ($t\approx 6$\,days), we find an upper limit magnitude consistent with this model.  

{\bf M15 precursor model:}  The photometry and spectroscopy acquired here show a high optical luminosity and hot, blue continua during the first $\sim$1\,d (see Sec.\,\ref{subsec: BB fits}).  Therefore, we explore the M15 model that predicts an energetic blue precursor. This model is based on the idea that a small fraction (i.e. $M_n \sim 10^{-4} M_\odot$ \citep{Metzger2017b}) of the ejected mass in the outer shell is rapidly expanded after shock heating during the merger.  Thus, the neutrons in the outer shell avoid capture by the nuclei in the dense inner ejecta during the r-process.  The unbound neutrons are then subject to $\beta$-decay which gives rise to a precursor to the kilonova which, at the distance to AT2017gfo, would peak at mag$_r \sim 17.5$ after a few hours, and consistent with the photometry.  The peak luminosity of the neutron layer can be approximated by $L_{\mathrm{peak}} \propto \left(\frac{v_{ej}}{0.1c}\right) \left(\frac{M_{ej}}{10^{-2}\,M_{\odot}}\right)^{1/3}$ \citep{Metzger2017b}. We overlay the {\it gri}-bands plots to our data in Figure~\ref{fig:TanakaNSNS} for lanthanide-free ejecta and for two sets of values for the velocity and mass of the ejecta ($v_{ej}=0.2c$, $M_{ej}=0.01\,M_{\odot}$; and $v_{ej}=0.2c$, $M_{ej}=0.1\,M_{\odot}$). 

The M15 model seems to match our observations with a greater accuracy than the TH13 and BK13 models in the first $\sim$2\,d after the merger.  However, this model alone predicts a steeper decay of the light curve than the observations.  The SkyMapper $g$-band upper limits place a mild constraint in favor of a scenario with only an M13-type precursor.  Nevertheless, the combination of the M15 and BK13 models, represented with black dashed lines in Figure\,\ref{fig:TanakaNSNS}, is a better match to our data and, particularly, for the $r$-band measurements shown in the central panel.

\section{DISCUSSION}
\label{sec: conclusion}
The first detection of the EM counterpart to a GW event is a milestone in the history of modern astronomy.  Australian teams contributed to both the search and the follow up of the transient AT2017gfo, the EM counterpart to GW170817.  In this paper we present the observations, follow-up strategies, and data acquired by 14 radio, infrared, and optical facilities led by Australian observing programs. 

It is interesting to discuss the progenitor of this event.  Our own galaxy contains at least 7 binary neutron star pairs that will coalesce in less than a Hubble time, see \citep{Ozel2016}.  For some, like the double pulsar PSR J0737-3039A/B \citep{Burgay2003} the ``remaining time'' before merger is short ($\sim$80 Myr) whereas for others, like PSR B1534+12 the (remaining) coalescence time is 2.7 Gyr \citep{Arz1999}.  The latter would appear to be a more likely progenitor for this event as it could have formed when the last episode of star formation in NGC~4993 was still underway.  It will be fascinating to see how many binary star mergers are ultimately observed in active star-forming galaxies from ``ultra-relativistic'' progenitors with short lifetimes compared to those from wider systems like PSR B1534+12. 

The location in NGC~4993 is also of some interest.  At 40~Mpc, the projected distance of AT2017gfo from the centre is 2.2~kpc.  Such a displacement could be achieved during a galaxy merger, so constraints on any kick received by the binary are poor.

No radio source is detected down to 40\,$\upmu$Jy with the ATCA, ASKAP, VLBA, and MWA telescopes within 10 days from the GW detection.  However, past sGRBs that were detected in the radio despite being 30--60 times more distant than this event \citep{Berger2005,Soderberg2006,Fong2014,Zhang2017} imply that future neutron star mergers at these (40 Mpc) distances could reach flux densities of 0.1 to 1~Jy.  The Parkes and ASKAP radio telescopes searched for FRBs in NGC~4993 after the binary neutron stars merger for a total of 5 and 18.7 hours, respectively.  No FRB was detected: a signal from a source at $\sim 40$~Mpc with similar properties of the repeating FRB~121102 \citep{Spitler2016} would have resulted in a highly significant detection. 

We compared ``standard'' sGRB afterglow models (GS02) with the optical light curve obtained with the measurements of the Zadko, AST3-2, SkyMapper, and Etelman/VIRT telescopes.  The AT2017gfo transient was proven to be the electromagnetic counterpart to GW170817 and GRB~170817A \citep{LVCGW170817MMA}, but its optical light curve does not match the sGRB afterglow models. The continuum profiles and evolution of the spectra of AT2017gfo are unlike sGRBs and argue for a kilonova-like explosion, with a blackbody-like event cooling rapidly over the $\sim$7\,d of our spectral coverage.  We compared three kilonova models (T\&H13, B\&K13, M15) with our photometric data and the plots, combined with the spectral evolution of the transient, make the combination of a neutron-powered blue precursor and a r-process red emission at later time a plausible scenario.

\section{FUTURE PERSPECTIVES} Several facilities discussed here have existing reactive programs to follow up GW alerts, while others perform observations as part of DWF and/or OzGrav.  

DWF coordinates $\sim$30 major observatories worldwide and in space to provide simultaneous, fast-cadenced, deep (m $\sim$ 23--25, optical), radio to gamma-ray coverage of fast transients and GW events\footnote{\url{http://www.dwfprogram.altervista.org}} \citep{Cooke2017}.  As a result, DWF is on-source before, during, and after fast transients and has been in full operation since 2016.  Moreover, DWF performs real-time (seconds) supercomputing data analysis and transient identification \citep{Andreoni2017mary, Vohl2017,Meade2017} and triggers rapid-response, conventional ToO, and long-term spectroscopy and imaging with our network of 1--10m-class telescopes.  DWF operates several weeks a year and was not on sky during this GW event.  However, 10 DWF participating facilities provided data for AT2017gfo.  GW event detections during future DWF observing runs will provide complete, densely-sampled, multi-wavelength imaging and spectroscopy of the event and host galaxy.

The intent of the EM component of OzGrav is to help oversee a number of collaborating facilities, including the DWF program, in an effort to optimise the follow up of GW events by Australian and Australian-led programs at all wavelengths.  By the time of LIGO/Virgo `O3' run, OzGrav will be fully optimised to provide complete and dense coverage of GW events at all wavelengths via imaging, spectroscopy, interferometry and fast radio burst searches.  

The DFN is being augmented with cameras designed to detect bright optical transients.  The first such station consists of a Nikon D810 Camera with a Samyang 14mm f/2.8 IF ED UMC Lens, giving a field of view of 80x100\,deg, an imaging cadence of 15~s, and a limiting magnitude of mag$_{v}$=10.  Successor astronomy stations have been developed to have a greater sky coverage and increased sensitivity, via multiple cameras to tile the sky with a $<$5~s imaging cadence and limiting magnitude of ag$_{v}$=12.  The current and future DFN network is the only facility that can provide continuous monitoring for half of the Southern sky.

The future of the OzGrav facilities network also includes the Gravitational-wave Optical Transient Observer (GOTO\footnote{\url{http://goto-observatory.org}}), a planned wide-field robotic optical telescope optimised for following up LVC triggers.  GOTO is supported by a collaboration between Monash University; Warwick, Sheffield, Leicester and Armagh University in the UK; and the National Astronomical Research Institute of Thailand (NARIT).  Each instrument consists of eight 40-cm astrographs on a single mount, with fields-of-view arranged to achieve a total coverage of order 40~square~degrees.  The prototype instrument, with four astrographs, was deployed in 2017 June, although full robotic operation was not achieved before the end of O2.  Funding has now been secured for an additional four astrographs, and the instrument is expected to commence operations in 2018.

Australia will further be able to support the search for and characterisation of GW sources with GLUV, a 30cm ultraviolet survey telescope under development at ANU \citep{Sharp2016} for a high altitude balloon platform. It will feature a 7~deg$^2$ field of view and a limiting magnitude in near-UV of $\sim$22. \cite{Ridden-Harper2017} explores the application of GLUV to gravitational wave source characterisation, showing that early UV observations could provide a powerful diagnostic to identify merger pathways. The system is expected to fly in 2019 and build towards a constellation of telescopes flying in observation campaigns.

\section{Acknowledgements}
We thank Prof. Brian Metzger who provided the {\it gri} light curves for the M15 model.  Part of this research was funded by the Australian Research Council Centre of Excellence for Gravitational Wave Discovery (OzGrav), CE170100004 and the Australian Research Council Centre of Excellence for All-sky Astrophysics (CAASTRO), CE110001020.   Parts of this research were conducted by the Australian Research Council Centre of Excellence for All-sky Astrophysics in 3 Dimensions (ASTRO-3D), CE170100013.

Research support to IA is provided by the Australian Astronomical Observatory (AAO).
JC acknowledges the Australian Research Council Future Fellowship grant FT130101219.  
The Etelman Observatory team acknowledges support through NASA grants NNX13AD28A and NNX15AP95A.
TM acknowledges the support of the Australian Research Council through grant FT150100099.
SO acknowledges the Australian Research Council grant Laureate Fellowship FL15010014.
DLK and ISB are additionally supported by NSF grant AST-141242.1  
PAB and the DFN team acknowledge the Australian Research Council for support under their Australian Laureate Fellowship scheme. C.M. is supported by NSF grant AST-1313484.

The Australia Telescope Compact Array is part of the Australia Telescope National Facility which is funded by the Australian Government for operation as a National Facility managed by CSIRO.

This scientific work makes use of the Murchison Radio-astronomy Observatory, operated by CSIRO. We acknowledge the Wajarri Yamatji people as the traditional owners of the Observatory site. Support for the operation of the MWA is provided by the Australian Government (NCRIS), under a contract to Curtin University administered by Astronomy Australia Limited. We acknowledge the Pawsey Supercomputing Centre which is supported by the Western Australian and Australian Governments.

The Australian SKA Pathfinder is part of the Australia Telescope National Facility which is managed by CSIRO. Operation of ASKAP is funded by the Australian Government with support from the National Collaborative Research Infrastructure Strategy. ASKAP uses the resources of the Pawsey Supercomputing Centre. Establishment of ASKAP, the Murchison Radio-astronomy Observatory and the Pawsey Supercomputing Centre are initiatives of the Australian Government, with support from the Government of Western Australia and the Science and Industry Endowment Fund. We acknowledge the Wajarri Yamatji people as the traditional owners of the Observatory site.  This work was supported by resources provided by the Pawsey Supercomputing Centre with funding from the Australian Government and the Government of Western Australia.

The Long Baseline Observatory is a facility of the National Science Foundation operated under cooperative agreement by Associated Universities, Inc.

The Zadko Telescope is supported by the University of Western Australia Department of Physics, in the Faculty of Engineering and Mathematical Sciences. We also thank the superb technical support from J. Moore and A. Burrel that has enabled the facility to participate in this project.

SkyMapper is owned and operated by The Australian National University's Research School of Astronomy and Astrophysics. The national facility capability for SkyMapper has been funded through ARC LIEF grant LE130100104 from the Australian Research Council, awarded to the University of Sydney, the Australian National University, Swinburne University of Technology, the University of Queensland, the University of Western Australia, the University of Melbourne, Curtin University of Technology, Monash University and the Australian Astronomical Observatory.

The AST3 project is supported by the National Basic Research Program (973 Program) of China (Grant No. 2013CB834900), and the Chinese Polar Environment Comprehensive Investigation \& Assessment Program (Grand No. CHINARE2016-02-03-05). The construction of the AST3 telescopes has received fundings from Tsinghua University, Nanjing University, Beijing Normal University, University of New South Wales, Texas A\&M University, the Australian Antarctic Division, and the National Collaborative Research Infrastructure Strategy (NCRIS) of Australia. It has also received fundings from Chinese Academy of Sciences through the Center for Astronomical Mega-Science and National Astronomical Observatory of China (NAOC). 

Based in part on data acquired through the Australian Astronomical Observatory. We acknowledge the traditional owners of the land on which the AAT stands, the Gamilaraay people, and pay our respects to elders past and present.

Some of the observations reported in this paper were obtained with the Southern African Large Telescope (SALT) under the Director's Discretionary Time programme 2017-1-DDT-009.
The SALT/SAAO team are supported by the National Research Foundation (NRF) of South Africa. 

Research partially based on observations collected at the European Organisation for Astronomical Research in the Southern Hemisphere under ESO programme 60.A-9392.


\clearpage

\begin{table*}[ht]
\centering
\caption{AST3-2 Observations of GW170817 and AT2017gfo}
\label{table: obs and photometry AST3}
\begin{tabular}{llll}
\hline \hline
\multicolumn{4}{c}{AST3-2}\\
UT obs date         & Band & Mag   & Mag. error  \\ 
\hline
2017-08-18 13:11:42.72 & i & 17.23 & 0.22(-0.21)  \\
2017-08-18 14:15:54.29 & i    & 17.61 & 0.16      \\
2017-08-18 15:00:16.24 & i    & 17.72 & 0.18(-0.17)    \\
2017-08-20 16:07:27.71 & i    & $>$18.67 &       \\
2017-08-21 15:36:49.65 & i    & $>$18.38 &       \\
\hline
\end{tabular}
\label{tab:ast}
\end{table*}

\begin{table*}[ht]
\centering
\caption{Zadko Observations of GW170817 and AT2017gfo}
\label{table: obs and photometry Zadko}
\begin{tabular}{llll}
\hline \hline
\multicolumn{4}{c}{Zadko}\\
UT obs date         & Band & Mag   & Mag. error  \\ 
\hline
2017-08-19 10:57:00 & r    & 18.46 & 0.17        \\
2017-08-20 11:30:00 & r    & 19.18 & 0.12        \\
2017-08-21 11:52:00 & r    & 19.86 & 0.21        \\
2017-08-22 11:46:00 & r    & 20.20  & 0.23       \\
2017-08-23 11:32:00 & r    & $>$20.6  &          \\
\hline
\end{tabular}
\end{table*}

\begin{table*}[ht]
\centering
\caption{Etelman/VIRT Observations of GW170817 and AT2017gfo.}
\label{table: obs and photometry VIRT}
\begin{tabular}{llll}
\hline \hline
\multicolumn{4}{c}{Etelman/VIRT}\\
UT obs date       & Band    & Mag      & Mag error\\ 
\hline
 2017-08-20 00:12 & Clear   & 18.90    & 0.28 \\   
 \hline
 \end{tabular}
 \end{table*}

\begin{table*}[ht]
\centering
\caption{ESO VLT Observations of GW170817 and AT2017gfo.  Kasliwal et al. 2017, in preparation}
\label{table: obs and photometry VLT}
\begin{tabular}{llll}
\hline \hline
\multicolumn{4}{c}{ESO VLT}\\
Instrument & UT obs date         & Band & Mag   \\ 
\hline
NACO &  2017-08-25 22:45  &  $L'3.8$ &  $>14.5$ \\
NACO &  2017-08-26 22:45  &  $L'3.8$ &  $>14.8$ \\
NACO &  2017-08-27 22:45  &  $L'3.8$ &  $>14.5$ \\
NACO &  2017-09-01 22:45  &  $L'3.8$ &  $>14.3$ \\
VISIR &  2017-08-23 23:35  &  $J8.9$ &  $>8.26$ \\
VISIR &  2017-08-31 23:18  &  $J8.9$ &  $>7.74$ \\
VISIR &  2017-09-01 23:18  &  $J8.9$ &  $>7.57$ \\
VISIR &  2017-09-06 23:33  &  $J8.9$ &  $>7.42$ \\
\hline
\end{tabular}
\end{table*}


\onecolumn
\begin{longtable}{llll}
\caption[]{SkyMapper Observations of GW170817 and AT2017gfo with photometric measurements. The SkyMapper follow up is not limited to the data points presented in this table. The results from the analysis of the complete dataset will be discussed in future publications. }\tabularnewline
\hline \hline
\multicolumn{4}{c}{SkyMapper}\\
UT obs date         & Band & Mag   & Mag. error  \\ 
\hline
2017-08-18 09:16:58 & i &   17.42 &    0.05 \\
2017-08-18 10:03:44 & r &   17.32 &    0.07 \\
2017-08-18 10:05:44 & g &   17.46 &    0.08 \\
2017-08-19 09:06:17 & i &   17.96 &    0.07 \\
2017-08-19 09:24:57 & i &   18.18 &    0.08 \\
2017-08-20 09:12:57 & r &   19.34 &    0.08 \\
2017-08-20 09:14:58 & g &   20.43 &    0.11 \\
2017-08-20 09:31:44 & r &   19.37 &    0.09 \\
2017-08-20 09:33:45 & g &   20.21 &    0.12 \\
2017-08-22 09:09:22 & r & $>$ 20.51 &    95\% \\
2017-08-22 09:11:24 & g & $>$ 20.60 &    95\% \\
2017-08-22 09:28:08 & r & $>$ 20.47 &    95\% \\
2017-08-22 09:30:08 & g & $>$ 20.66 &    95\% \\
2017-08-28 09:17:13 & r & $>$ 19.36 &    95\% \\
2017-08-28 09:19:13 & g & $>$ 19.53 &    95\% \\
2017-08-28 09:35:52 & r & $>$ 19.39 &    95\% \\
2017-08-28 09:37:53 & g & $>$ 19.50 &    95\% \\
2017-08-30 09:18:53 & g & $>$ 19.36 &    95\% \\
2017-08-30 09:20:52 & r & $>$ 19.32 &    95\% \\
2017-08-30 09:37:33 & g & $>$ 19.24 &    95\% \\
 \hline
\label{table: obs and photometry SkyMapper}
\end{longtable}
\nopagebreak
\twocolumn
\nopagebreak


\begin{table*}[ht]
\centering
\caption{ASKAP Observations of GW170817 and AT2017gfo}
\label{table: obs and photometry ASKAP}
\begin{tabular}{llcccc}
\hline \hline
\multicolumn{6}{c}{ASKAP}\\
UT obs date & Mode & Frequency (MHz) & Bandwidth (MHz) & N$_{\textrm{ant.}}$ & N$_{\textrm{beams}}$\\
\hline
2017-08-18 04:05--07:36 & FRB &  1320 & 336 & 7 & 108 \\
2017-08-18 08:57--13:03 & FRB &  1320 & 336 & 7 & 108 \\
2017-08-19 02:08--13:08 & FRB &  1320 & 336 & 7 & 108 \\
2017-08-19 05:34--07:58 & Imaging &  1344 & 192 & 10 & 36\\
2017-08-20 02:21--11:21 & Imaging &  1344 & 192 & 10 & 36\\
2017-08-21 07:21--12:28 & Imaging &  1344 & 192 & 10 & 36\\
2017-08-22 01:44--10:52 & Imaging &  1344 & 192 & 10 & 36\\
2017-09-01 02:33--03:28 & Imaging &  888 & 192 & 12 & 1\\
2017-09-01 07:59--10:59 & Imaging &  888 & 192 & 12 & 1\\
2017-09-02 06:21--08:28 & Imaging &  888 & 192 & 16 & 1\\
2017-09-06 01:16--02:17 & Imaging &  1344 & 192 & 12 & 1\\
2017-09-06 03:36--08:36 & Imaging &  1344 & 192 & 12 & 1\\
2017-09-08 02:32--06:00 & Imaging &  1344 & 192 & 16 & 1\\
2017-09-09 03:34--08:41 & Imaging &  1344 & 192 & 16 & 1\\
2017-09-10 03:52--04:52 & Imaging &  1344 & 192 & 16 & 1\\
2017-09-15 08:17--11:17 & Imaging &  1344 & 192 & 15 & 1\\
2017-09-21 05:30--06:30 & Imaging &  1344 & 192 & 12 & 1\\
2017-09-22 08:35--10:35 & Imaging &  1368 & 240 & 12 & 1\\
2017-09-29 23:21--2017-09-30 03:21 & Imaging &  1320 & 240 & 12 & 36\\
2017-09-30 23:32--2017-10-01 03:32 & Imaging &  1320 & 240 & 12 & 36\\
2017-10-01 23:32--2017-10-02 03:32 & Imaging &  1320 & 240 & 12 & 36\\
\hline
\end{tabular}
\end{table*}

\begin{table*}[ht]
\centering
\caption{ATCA Observations of GW170817 and AT2017gfo.}
\label{table: obs and photometry ATCA}
\begin{tabular}{lccc}
\hline \hline
\multicolumn{4}{c}{ATCA (Imaging)}\\
UT obs date  &  Frequency (GHz) & Bandwidth (GHz) & Flux ($\upmu$Jy) \\ 
\hline
2017-08-18 01:00--09:07  &  8.5 & 2.049 & $<$ 120 \\
2017-08-18 01:00--09:07  &  10.5 & 2.049 & $<$ 150  \\
2017-08-18 01:00--09:07  &  16.7 & 2.049 & $<$ 130  \\
2017-08-18 01:00--09:07  &  21.2 & 2.049 & $<$ 140  \\
2017-08-20 23:31--2017-08-21 11:16  &  8.5 & 2.049 & $<$ 135  \\
2017-08-20 23:31--2017-08-21 11:16  &  10.5 & 2.049 & $<$ 99  \\
2017-08-27 23:31--2017-08-28 09:00  &  8.5 & 2.049 & $<$ 54  \\
2017-08-27 23:31--2017-08-28 09:00  &  10.5 & 2.049 & $<$ 39  \\
2017-08-27 23:31--2017-08-28 09:00  &  10.5 & 2.049 & $<$ 39  \\
2017-09-04 22:48--2017-09-05 10:04  &  7.25 & 4.098 & 25$\pm$6 \\
\hline
\end{tabular}
\end{table*}

\begin{table*}
 \centering
  \caption{ATCA measured flux densities for NGC 4993}
  \label{tab:ATCANGC}
  \begin{tabular}{lcc}
    \hline
    \hline 
   Observation Date  & Frequency & Flux Density\\
     (UTC) & (GHz) & ($\upmu$Jy) \\
    \hline
    2017-08-18.04\,--\,2017-08-18.38 & 8.5  & 420$\pm$50 \\
    2017-08-20.98\,--\,2017-08-21.47 & 8.5  & 360$\pm$20 \\
    2017-08-27.98\,--\,2017-08-28.37 & 8.5  & 460$\pm$30 \\
    2017-08-18.04\,--\,2017-08-18.38 & 10.5 & 500$\pm$40 \\
    2017-08-20.98\,--\,2017-08-21.47 & 10.5 & 550$\pm$60 \\
    2017-08-27.98\,--\,2017-08-28.37 & 10.5 & 400$\pm$20 \\
    2017-08-18.04\,--\,2017-08-18.38 & 16.7 & 300$\pm$50 \\
    2017-08-18.04\,--\,2017-08-18.38 & 21.2 & 210$\pm$70 \\
    \hline
  \end{tabular}
 \end{table*}

\begin{table*}[ht]
\centering
\caption{MWA Observations of GW170817 and AT2017gfo}
\label{table: obs and photometry MWA}
\begin{tabular}{lcc}
\hline \hline
\multicolumn{3}{c}{MWA (Imaging)}\\
 UT obs date  & Frequency (MHz) & Bandwidth (MHz) \\ 
\hline
2017-08-18 07:07:52--09:40:00  & 185 & 30.72 \\
2017-08-19 07:04:00--09:38:00   & 185 & 30.72\\
2017-08-20 07:00:08--09:34:08   & 185 & 30.72\\
2017-08-21 06:56:08--09:28:08   & 185 & 30.72\\
2017-08-22 06:52:16--09:26:16   & 185 & 30.72\\
\hline
\end{tabular}
\end{table*}

\begin{table*}[ht]
\centering
\caption{Parkes Observations of GW170817 and AT2017gfo searching for FRBs.}
\label{table: obs and photometry Parkes}
\begin{tabular}{lcc}
\hline \hline
\multicolumn{3}{c}{Parkes (FRB)}\\
UT obs date   & Frequency (GHz) & Bandwidth (MHz) \\ 
\hline
2017-08-18 06:49:31  & 1.341 & 340 \\
2017-08-18 08:50:36  & 1.341 & 340 \\
2017-08-20 01:44:32  & 1.341 & 340 \\
2017-08-20 02:50:14  & 1.341 & 340 \\
\hline
\end{tabular}
\end{table*}

\begin{table*}[ht]
\centering
\caption{VLBA Observations of GW170817 and AT2017gfo}
\label{table: obs and photometry VLBA}
\begin{tabular}{lccc}
\hline \hline
\multicolumn{4}{c}{VLBA (Imaging)}\\
UT obs date   & Frequency (GHz) & Bandwidth (MHz) & Flux ($\upmu$Jy)\\ 
\hline
2017-08-18 19:58 -- 2017-08-19 01:34  & 8.7 & 256 & $<$ 125\\
2017-08-20 18:31 -- 2017-08-21 01:13  & 8.7 & 256 & $<$ 125\\
2017-08-21 18:26 -- 2017-08-22 01:08  & 8.7 & 256 & $<$ 120\\
\hline
\end{tabular}
\end{table*}

\begin{table*}[ht]
\centering
\caption{ANU2.3/WiFeS Observations of GW170817 and AT2017gfo}
\label{table: obs and photometry ANU2.3}
\begin{tabular}{lll}
\hline \hline
\multicolumn{3}{c}{ANU2.3/WiFeS}\\
UT obs date  & Spectral Range (\AA) &  Exposure (s)  \\
\hline
 2017-08-18 09:24:25 & 3300-9800 & 900  \\
2017-08-18 09:40:25 & 3300-9800 & 900  \\
2017-08-19 08:43:15 & 3300-9800 & 900  \\
2017-08-19 08:59:42 & 3300-9800 & 900  \\
2017-08-19 09:16:06 & 3300-9800 & 900  \\
2017-08-19 09:36:18 & 3300-9800 & 900  \\
2017-08-19 09:55:38 & 3300-9800 &  900 \\
2017-08-20 08:47:28 & 3200-7060 &1800   \\
2017-08-20 09:21:33 & 3300-9800 & 1800  \\
2017-08-21 08:40:58 & 3300-9800 &  900  \\
2017-08-21 09:13 & 3300-9800 & 900  \\
2017-08-21 09:29 & 3300-9800 & 900  \\
\end{tabular}
\end{table*}

\begin{table*}[ht]
\centering
\caption{SALT/RSS Observations of GW170817 and AT2017gfo}
\label{table: obs and photometry SALT}
\begin{tabular}{lll}
\hline \hline
\multicolumn{3}{c}{SALT/RSS}\\
UT obs date  & Spectral Range (\AA) &  Exposure (s)  \\
\hline
2017-08-18 17:07:19.703 & 3600-8000 & 433 \\
2017-08-19 16:58:32.76 & 3600-8000 & 716 \\
\end{tabular}
\end{table*}

\begin{table*}[ht]
\centering
\caption{AAT/AAOmega+2dF Observations of GW170817 and AT2017gfo}
\label{table: obs and photometry AAT/AAO}
\begin{tabular}{lll}
\hline \hline
\multicolumn{3}{c}{AAT/AAOmega+2dF}\\
UT obs date  & Spectral Range (\AA) &  Exposure (s)  \\
\hline
2017-08-24 08:55:07 &3750-8900 &  2400 \\
\hline
\end{tabular}
\end{table*}

\clearpage
\bibliographystyle{pasa-mnras}

\bibliography{references_EM,gcn}

\clearpage

\end{document}